\documentclass{aa}  

\usepackage{graphicx}
\usepackage{txfonts}
\usepackage[breaklinks=true]{hyperref}
\usepackage{ctable}
\usepackage{mathtools}
\usepackage{natbib,twoopt}
\usepackage[breaklinks=true]{hyperref} %% to avoid \citeads line fills
\hypersetup{colorlinks=true, linkcolor=blue, urlcolor=blue, citecolor=blue}
\bibpunct{(}{)}{;}{a}{}{,} %% natbib format for A&A and ApJ
\makeatletter
\newcommandtwoopt{\citeads}[3][][]{\href{http://adsabs.harvard.edu/abs/#3}%
{\def\hyper@linkstart##1##2{}%
\let\hyper@linkend\@empty\citealp[#1][#2]{#3}}}
\newcommandtwoopt{\citepads}[3][][]{\href{http://adsabs.harvard.edu/abs/#3}%
{\def\hyper@linkstart##1##2{}%
\let\hyper@linkend\@empty\citep[#1][#2]{#3}}}
\newcommandtwoopt{\citetads}[3][][]{\href{http://adsabs.harvard.edu/abs/#3}%
{\def\hyper@linkstart##1##2{}%
\let\hyper@linkend\@empty\citet[#1][#2]{#3}}}
\newcommandtwoopt{\citeyearads}[3][][]%
{\href{http://adsabs.harvard.edu/abs/#3}
{\def\hyper@linkstart##1##2{}%
\let\hyper@linkend\@empty\citeyear[#1][#2]{#3}}}
\makeatother

\newcommand{\dg}{$^{\circ}$}

\newcommand{\adeg}[1]{{#1}$^{\circ}$}
\newcommand{\amin}[1]{{#1}$^\prime$}
\newcommand{\asec}[1]{{#1}$^{\prime\prime}$}

\newcommand{\mjy}[1]{{#1}\,mJy}
\newcommand{\mjybeam}[1]{{#1}\,mJy\,beam$^{-1}$}

\newcommand{\ujybeam}[1]{{#1}\,$\mu$Jy\,beam$^{-1}$}

\definecolor{mygreen}{rgb}{0.0, 0.42, 0.0}

\usepackage[mathlines,switch]{lineno}
\newcommand{\nsources}{249672~}
\newcommand{\aplova}{22825~}
\newcommand{\aplo}{152824~}

\newcommand{\apnvss}{44523~}

\begin{document} 

   \title{Continuum source catalog for the first APERTIF data release}

   \author{
            A.\,M.\,Kutkin
            \inst{1}
        \and
            T.\,A.\,Oosterloo
            \inst{1,2}
        \and
            R.\,Morganti
            \inst{1,2}
        \and
            E.\,A.\,K.\,Adams
            \inst{1,2}
        \and
            M.\,Mancini
            \inst{1}
        \and
            B.\,Adebahr
            \inst{3}
        \and
            W.\,J.\,G.\,de\,Blok
            \inst{1,4,2}
        \and
            H.\,D\'{e}nes
            \inst{1}
        \and
            K.\,M.\,Hess
            \inst{5,1,2}
        \and
            J.\,M.\,van\,der\,Hulst
            \inst{2}            
        \and
            D.\,M.\,Lucero
            \inst{10}            
        \and
            V.\,A.\,Moss
            \inst{6,7,1}
        \and
            A.\,Berger
            \inst{3}
        \and
            R.\,van\,den\,Brink
            \inst{1}
        \and
            W.\,A.\,van\,Cappellen
            \inst{1}
        \and
            L.\,Connor
            \inst{8,1,9}
        \and
            S.\,Damstra
            \inst{1}
        \and
            G.\,M.\,Loose
            \inst{1}
        \and
            J.\,van\,Leeuwen
            \inst{1}
        \and
            Y.\,Maan
            \inst{13,1}
        \and
            A'.\,Mika
            \inst{1}
        \and
            M.\,J.\,Norden
            \inst{1}
        \and
            A.\,R.\,Offringa
            \inst{1,2}
        \and
            L.\,C.\,Oostrum
            \inst{1,8,11}
        \and
            D.\,van\,der\,Schuur
            \inst{1}
        \and
            D.\,Vohl
            \inst{8,1}
        \and
            S.\,J.\,Wijnholds
            \inst{1}
        \and
            J.\,Ziemke
            \inst{1,12}
        }
          
    \institute{
        ASTRON, The Netherlands Institute for Radio Astronomy, Oude Hoogeveensedijk 4, 7991 PD, Dwingeloo, The Netherlands\\
        \email{kutkin@astron.nl}
    \and
        Kapteyn Astronomical Institute, PO Box 800, 9700 AV Groningen, The Netherlands
    \and
        Astronomisches Institut der Ruhr-Universit{\"a}t Bochum (AIRUB), Universit{\"a}tsstrasse 150, 44780 Bochum, Germany
    \and
        Dept.\ of Astronomy, Univ.\ of Cape Town, Private Bag X3, Rondebosch 7701, South Africa        
    \and
        Instituto de Astrof\'{i}sica de Andaluc\'{i}a (CSIC), Glorieta de la Astronom\'{i}a s/n, 18008 Granada, Spain
    \and
        CSIRO Astronomy and Space Science, Australia Telescope National Facility, PO Box 76, Epping NSW 1710, Australia
    \and
        Sydney Institute for Astronomy, School of Physics, University of Sydney, Sydney, New South Wales 2006, Australia
    \and
        Anton Pannekoek Institute, University of Amsterdam, Postbus 94249, 1090 GE Amsterdam, The Netherlands
    \and
        Cahill Center for Astronomy, California Institute of Technology, Pasadena, CA, USA
    \and
        Department of Physics, Virginia Polytechnic Institute and State University, 50 West Campus Drive, Blacksburg, VA 24061, USA
    \and
        Netherlands eScience Center, Science Park 140, 1098 XG, Amsterdam, The Netherlands
    \and
        University of Oslo Center for Information Technology, P.O. Box 1059, 0316 Oslo, Norway
    \and
        National Centre for Radio Astrophysics, Tata Institute of Fundamental Research, Pune 411007, Maharashtra, India
    }

% \abstract{}{}{}{}{} 
% 5 {} token are mandatory
 
  \abstract
  {
  %context
  The first data release of Apertif survey contains 3074 radio continuum  images covering a thousand square degrees of the sky. The observations were performed during August 2019 to July 2020. The continuum images were produced at a central frequency 1355\,MHz with the bandwidth of $\sim 150$\,MHz and angular resolution reaching \asec{10}.
  %of $\sim 12\times12/sin(\mathrm{Declination})$ arcseconds.
  In this work we introduce and apply a new method to obtain a primary beam model using a machine learning approach, Gaussian process regression. The primary beam models obtained with this method are published along with the data products for the first Apertif data release. We apply the method to the continuum images, mosaic them and extract the source catalog.
%   We apply this correction to every image and its nearest overlapping neighbours, convolve them to a common restoring beam, re-project them on a common sky grid and produce a mosaic image with higher sensitivity. After extracting the sources in all the mosaic images and removing the duplicates we compile a full catalog. 
  The catalog contains \nsources radio sources many of which are detected for the first time at these frequencies. 
  We cross-match the coordinates with the NVSS, LOFAR/DR1/value-added and LOFAR/DR2 catalogs resulting in \apnvss, \aplova and \aplo common sources respectively. The first sample provides a unique opportunity to detect long term transient sources which have significantly changed their flux density for the last 25 years. The second and the third ones combined together provide information about spectral properties of the sources as well as the redshift estimates. 
  }

%TODO: check:

   \maketitle

\section{Introduction}

Large astronomical surveys have always had a key role in helping characterize and understand the objects which populate our Universe. Continuous technological developments facilitate deep observations with a large field of view, boosting even more the possibilities for producing new, large and deep surveys.
At radio frequencies, this growth in the number of  surveys is particularly evident, triggered by the plethora of new instruments and telescopes developed to prepare for the upcoming Square Kilometer Array (SKA, e.g., \citealt{2019arXiv191212699B}).
Importantly, in the great majority of the cases, 
the related source catalogs are made available online, providing the  astronomical community with invaluable tools for carrying out science not possible so far.
%Large and deep sky surveys are becoming more and more important for modern astrophysics, providing wonderful statistical data samples and challenging current computational facilities. Today surveys covering different frequency and spatial scales, complementary to each other, provide a great outcome. 

Among the SKA-pathfinders which have started producing new surveys, is Apertif (AperTIF -- Aperture Tile In Focus), the phased array feed (PAF) receiver working at 1400 MHz and installed on 12 of the dishes of the Westerbork Synthesis Radio Telescope (WSRT).
It provides forty instantaneous overlapping beams (in the case of a phased array feed also called compound beams, CB) creating a wide field of view of about 10 square degrees. This along with a wide bandwidth of 300\,MHz makes Apertif an excellent instrument for surveys. 
A backend with high spectral resolution (12.2\,kHz) provides the possibility for simultaneous spectral line, continuum and polarization surveys. The full system and its capabilities are described in~\citet{2022A&A...658A.146V}.

Apertif not only represents an excellent proof of concept for the future generation of phased array telescopes, it also provides valuable scientific data in terms of sensitivity and angular resolution and is a significant improvement over the the two major surveys of the northern sky at 20 cm  which were performed more than 20 years ago: the NRAO VLA Sky Survey~\citepads[NVSS;]{1998AJ....115.1693C} and VLA Faint Images of the Radio Sky at Twenty Centimeters (FIRST; \citealt{Becker95}). The former had a good surface brightness sensitivity of \mjybeam{0.45}, but a low angular resolution of \asec{45}, the latter was even more sensitive (\mjybeam{0.15}), but resolved out extended structure of many radio sources with its \asec{5} beam size. The Apertif survey provides an intermediate angular resolution of $\sim 12^{\prime\prime}\times12^{\prime\prime}/sin(\mathrm{Declination})$ together with a noise level reaching \ujybeam{20} in its most sensitive single-pointing images. Overlap with the above-mentioned surveys makes the Apertif continuum data a great platform for a plethora of scientific questions. In particular, a separation in time between the Apertif and NVSS surveys of about 25 years provides an opportunity for detection of long-term variable radio sources. 

Importantly, resolution and sensitivity of the Apertif survey are sufficient to complement the breakthrough LOFAR Two-meter Sky Survey (LoTSS; \citealt{2017A&A...598A.104S,2019A&A...622A...1S, 2022A&A...659A...1S}). With the LOFAR surveys aiming at covering the entire northern sky at 60 and 150\,MHz, there is a perfect synergy between them and the Apertif surveys. This has been already illustrated by the studies presented in, for example, \cite{2021Galax...9...88M, 2021A&A...648A...9M}. 

The potential of Apertif is further enhanced by the fact that the raw data as well as the data products are made publicly available to the scientific community. The data release of the first year of Apertif observations (Apertif Data Release 1, ADR1) took place in November 2020\footnote{\url{http://hdl.handle.net/21.12136/B014022C-978B-40F6-96C6-1A3B1F4A3DB0}} (Adams et al., submitted, A22 hereafter).
Processed data products for every CB were released on the basis of the continuum image quality, and all released data products include validation measures. In this paper we complement the continuum images with the extracted source catalog. 

A key requirement for a source catalog which can be used for science is to have a correct flux scale. For PAFs, it is a challenging task to measure the amplification diagram, a.k.a. the primary beam response. In Apertif, the 40 CBs are formed electronically, by combining and weighting the signals received by the antenna elements installed in the focal plane of every dish (Vivaldi antennas in the case of Apertif (see \citealt{2022A&A...658A.146V} for details). As a consequence, they have different shapes, which may change in time as well as per antenna. 
Recovering the CB shapes in a flexible, fast and reliable way is essential for PAF observations (see \citealt{2022arXiv220509662D}, for a discussion) to deliver reliable scientific products.

In this work we propose and apply a new method to recover a compound beam shape, based on the NVSS Survey and Gaussian process regression. 
We apply the method to the ADR1 continuum images, mosaic them and extract the source catalog. We also illustrate some potential for science that it is offering. 
The paper is organized as follows.
In section~\ref{sec:data} we describe the ADR1 data. In section~\ref{sec:cbeams} we introduce the ``primary beam'' correction.  The mosaicing procedure is described in section~\ref{sec:mosaicing}. The source extraction, compilation of the catalog, reliability, completeness, flux scale, and astrometry are discussed in section~\ref{sec:cataloging}. The results of cross-matching of the catalog with the LOFAR/LoTSS and the NVSS catalogs are presented in section~\ref{sec:science}.

%%%%%%%%%%%%%%%%%%%%%%%%%%%%%%%%%%%%%%%%%%%%%%%%%%%%%%%%%%%%

\section{The data}
\label{sec:data}

First scientific verification imaging observations with Apertif started in April 2019 followed by the launch of the full survey program on 1 July 2019. Apertif is undertaking a two-tiered imaging survey: a wide field single-pointing survey and a deeper survey with multiple pointings over a smaller area (Hess et al., in prep).

The first-year Apertif data release (ADR1) occurred in November 2020~(A22). This time period includes a total of 221 observations of 160 independent fields, covering about a thousand square degrees of sky. The released data products include continuum images, polarization images and cubes, and spectral line cubes, obtained with an automatic processing pipeline separately for every compound beam. 

\begin{figure}
    \centering
    \includegraphics[width=\columnwidth]{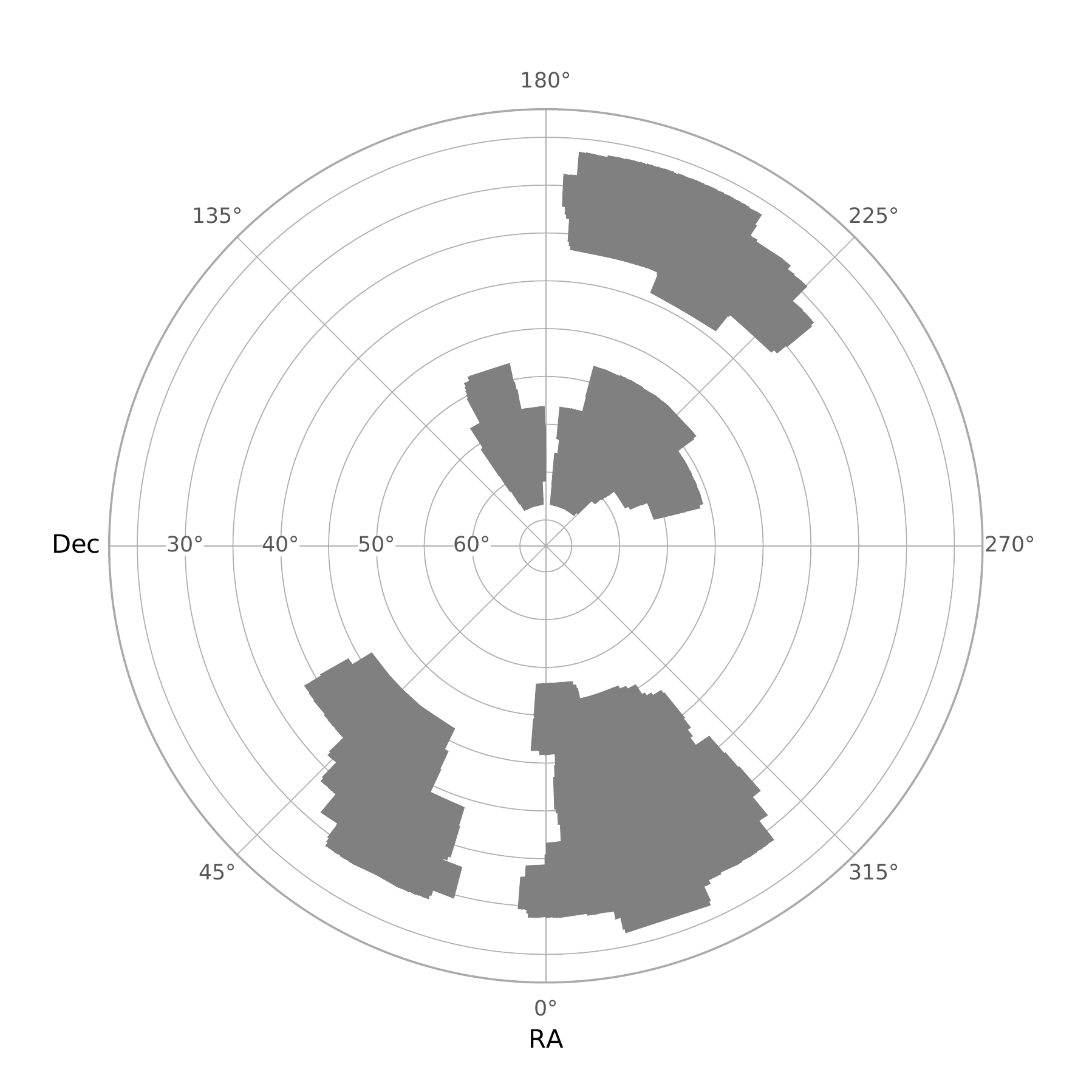}
    \caption{Sky coverage of the released continuum images.}
    \label{fig:sky_coverage}
\end{figure}

The WSRT array has east-west orientation and normally a field is being observed for 11--12\,hours using the Earth rotation to fill the $uv$-plane. Since the dishes have an equatorial mount, the CBs do not rotate or move with respect to the sky during an observation.
In a single observation, Apertif covers forty adjacent CBs with a total area of $\sim 10$ square degrees. The raw data for every CB is  processed independently with the automatic pipeline which includes flagging, calibration and imaging~\citep{2022A&C....3800514A}. For ADR1, the processed frequency range covers the 1292.5--1430 MHz interval. A dedicated validation procedure has been developed and applied based on inspection of the continuum images, which were required to have only modest calibration residuals and a noise level below \ujybeam{60}. Based on these criteria some CB images were rejected, which is mostly dictated by the limitations of the current software to handle direction-dependent effects. More details about the first data release of the Apertif imaging surveys are given by A22. In the end, a total of 3374 continuum Stokes I images of individual CBs have been released. The sky coverage of the released continuum images is shown in Figure~\ref{fig:sky_coverage}.

%If an image is too noisy or dominated by RFI/artifacts it is discarded from the release. For example, sources brighter that \mjy{100} located at a CB edge tend to have prominent artifacts in the images due to direction dependent effects, which are not handled by the current pipeline. Many of such images were then rejected by the validation procedure and were not released. This resulted in a corresponding lack of bright sources in the released images, which is seen at the differential source counts plot (Section~\ref{sec:completeness}). The validation procedure also causes a patchy sky coverage for processed data products in ADR1, which is to be improved in future data releases. 

Every image has a size of 3073 by 3073 pixels, with a pixel size of \asec{4} (the images are $200\times200\,'$ in size, much larger than a primary beam response in order to encompass side lobes for the cleaning procedure). 
In the Apertif field of view layout, the centers of the CBs are equally separated from each other by \amin{28}, resulting every CB to have up to six equidistant neighbors. The full width at half maximum (FWHM) of a CB radial attenuation profile  is about \amin{36}, however its shape may be significantly non-symmetric and changing in time (see below). Many of the CB images are obtained for the same sky position  in the framework of the medium-deep survey. 
The ADR1 data products are available through a virtual observatory interface\footnote{\url{https://vo.astron.nl/apertif_dr1/q/apertif_dr1_continuum_images/form}}. 

As already mentioned, for most scientific purposes the radial response of each CB needs to be characterized and corrected for. The primary beam characterization for the pre-Apertif WSRT was known very well \citep{2008A&A...479..903P}. However, the characteristics of the electronically formed beams from the Vivaldi antennas carry a number of complications~\citep{2022arXiv220509662D}. The corresponding CBs shapes differ from those of a single-receiver antenna and from each other, being more distorted towards the edges of the field of view. Moreover,  malfunctioning of any of the PAF elements, or a leakage between them, can cause a significant deformation of a CB shape on a given dish. As a result,  the primary beam correction used for the old WSRT is not applicable to the 40 CBs of Apertif. 

Below we introduce and implement an efficient approach for this correction. The new method has been applied to the ADR1 continuum images and the corresponding CB correction models were published along with with the processed data. 

%%%%%%%%%%%%%%%%%%%%%%%%%%%%%%%%%%%%%%%%%%%%%%%%%%%%%%%%%%%%

\section{Compound beam shapes}
\label{sec:cbeams}

Traditionally, holography is used to measure the primary beam response of radio telescopes; however Apertif did not support a holographic mode. An alternative approach is to perform drift scans of a bright calibrator, typically Cyg\,A or Cas\,A. 
However, this approach has several limitations. First, it can be affected by a complex source structure, background sources and/or radio frequency interference (RFI) present during scanning. Second, it is extremely time consuming. To perform one such calibration procedure for all 40 CBs requires more than 14 hours which is longer than a full synthesis observation of a survey pointing. Finally, these calibrators are extended sources resolved on WSRT baselines and their substructure complicates the analysis even for the auto-correlation amplitudes of the visibilities, while compact calibrators like 3C286 are not strong enough for drift-scanning. For this reason, such drift scans are performed only once per month or less frequently. Moreover, due to observing time limitations, the resulting models of a CB are often cut off well before the first null~\citep{2022arXiv220509662D}. While this approach is used in Apertif project providing a good reference, it is important to have an independent, faster and more flexible method to recover a CB shape for any given observation date.

The approach we propose and use here is based on the public NVSS catalog data combined with Gaussian process (GP) regression. The method takes advantage of the fact that NVSS survey is obtained at a very similar frequency as the Apertif survey, allowing one to determine the CB attenuation factor at any position in the field of view, for any given date of observation. 

A GP is a stochastic process with normally distributed values characterized by their mean and covariance function, or kernel. The kernel depends on hyperparameters and determines the characteristics of the GP. Data can be approximated using GP regression, providing a probabilistic prediction of the GP value at a given coordinate~\citep{Rasmussen:GPM}.

Although the NVSS was performed at a similar frequency, it has a lower angular resolution compared to Apertif. Therefore, the first step is to convolve the Apertif CB images with a circular PSF of \asec{45} to match the NVSS resolution. In the next step,  a source finder is used (as described in Section~\ref{sec:cataloging}) and sources are cross-matched with the NVSS catalog. Here we used the same approach for source extraction and cross identification as described below in Sections~\ref{sec:cataloging} and \ref{sec:flux_scale}. We use all the cross-matched sources to obtain the spatial distribution of the relation between NVSS and Apertif total flux, $\epsilon = S_\mathrm{APERTIF} / S_\mathrm{NVSS}$, over a given CB image. This relation, on average, represents the corresponding CB shape. 
 
%We note, however, there might be a minor bias due to a different sensitivity of the two surveys. Indeed, extended sources in the Apertif images tend to be ``more extended'' than because of better sensitivity and, therefore, may have higher integrated flux density on average. We checked that this effect is less than few percent within a CB clipped at level 0.1. (see /kutkin/gptest/beam01_compact_to_all.png)

The next step is to involve a GP and train it on the existing data. We select a  squared-exponential radial basis function for the kernel as the simplest way to describe smooth data variations. We construct the kernel as the sum of two squared-exponential kernels (SE1+SE2) and a white noise one: SE1 to represent the general CB shape, and SE2 to describe smaller-scale deformations related to e.g. individual PAF elements malfunctioning. For the training, we set a wide range of priors on the hyperparameters, and used all the data for a given CB consisting of a few thousand measurements (ADR1 implies about 40 images per CB). The hyperparameters were found to be plausible (the SE1 scale describes a CB width well and the SE2 amplitude is less than 10\% of the SE1 one) and stable with respect to varying sample size. After validation, we used the trained GP with the SE1 kernel to predict a CB shape. This allowed us to get the ``average'' CB shapes over the whole year period of the observations considered. The corresponding FITS images are published together with the ADR1 data and can be used to correct the continuum images. We note that the peak values of the CB radial profiles differ from 1 indicating potential differences in flux scale between Apertif and NVSS surveys. 

% Trained GP for beam01: 0.159**2 * RBF(length_scale=[348, 329]) + 0.0219**2 * RBF(length_scale=[13.5, 25.7]) + WhiteKernel(noise_level=1e-05)

\begin{figure*}
\begin{center}
    \includegraphics[width=\columnwidth]{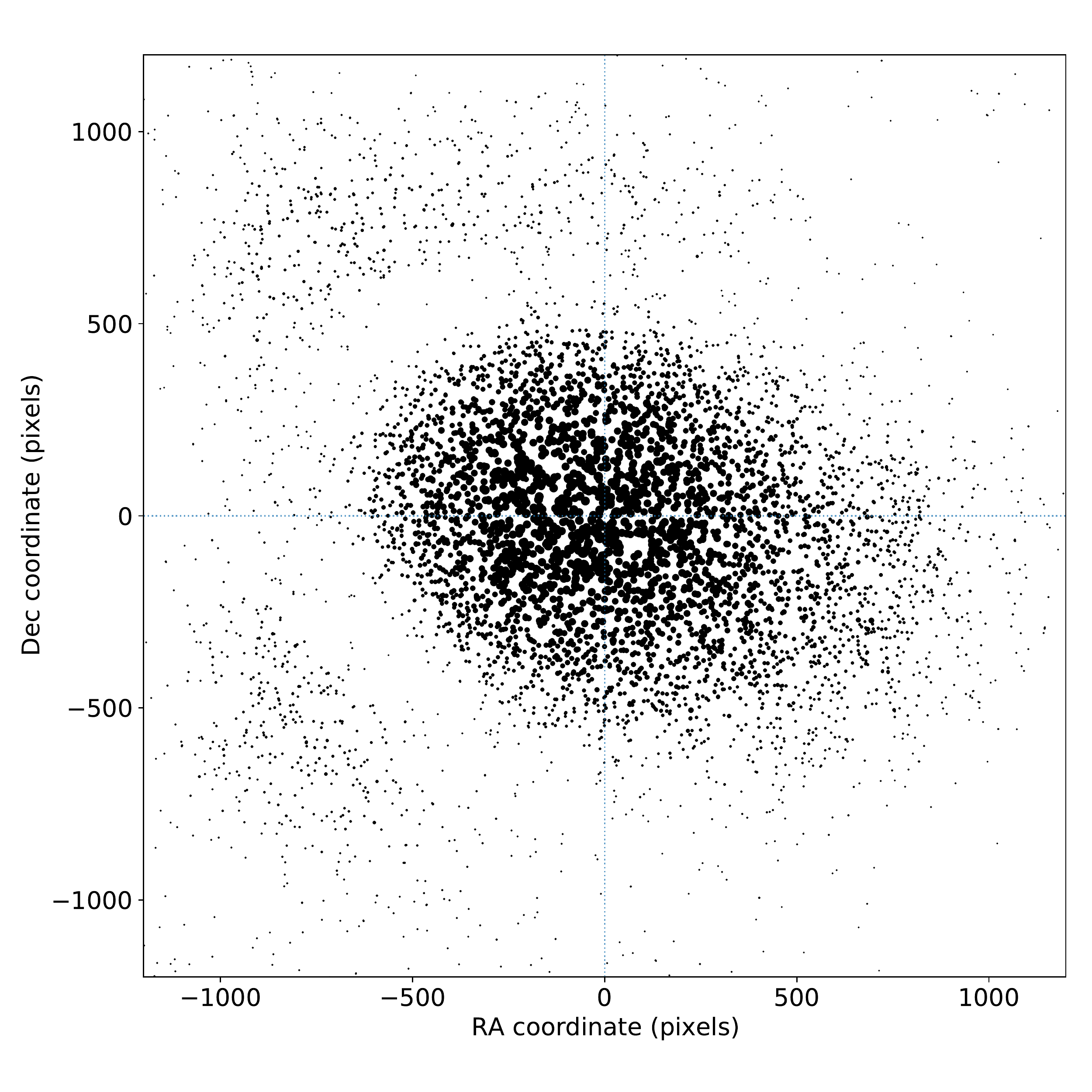}~\includegraphics[width=\columnwidth]{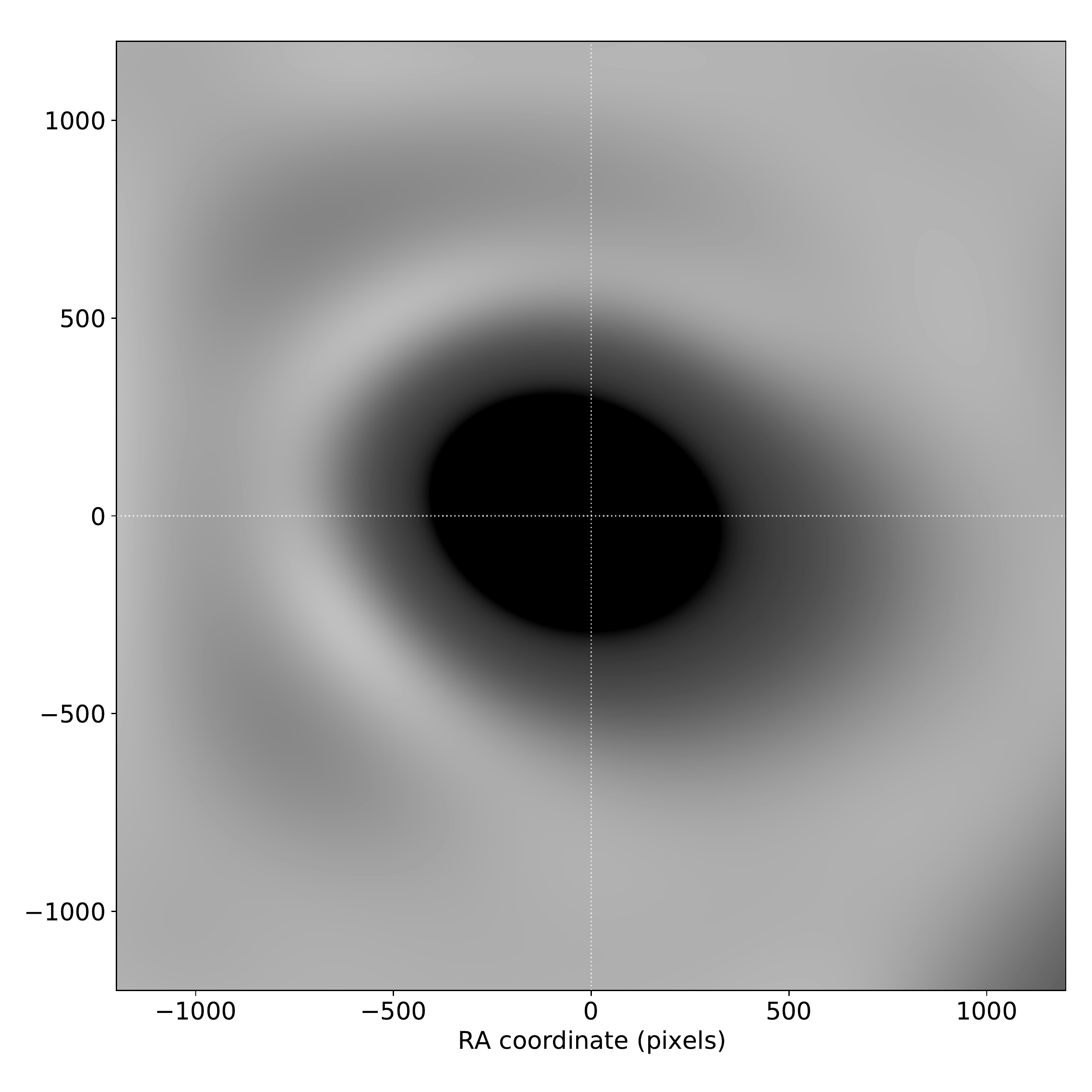}\\
    \includegraphics[width=\columnwidth]{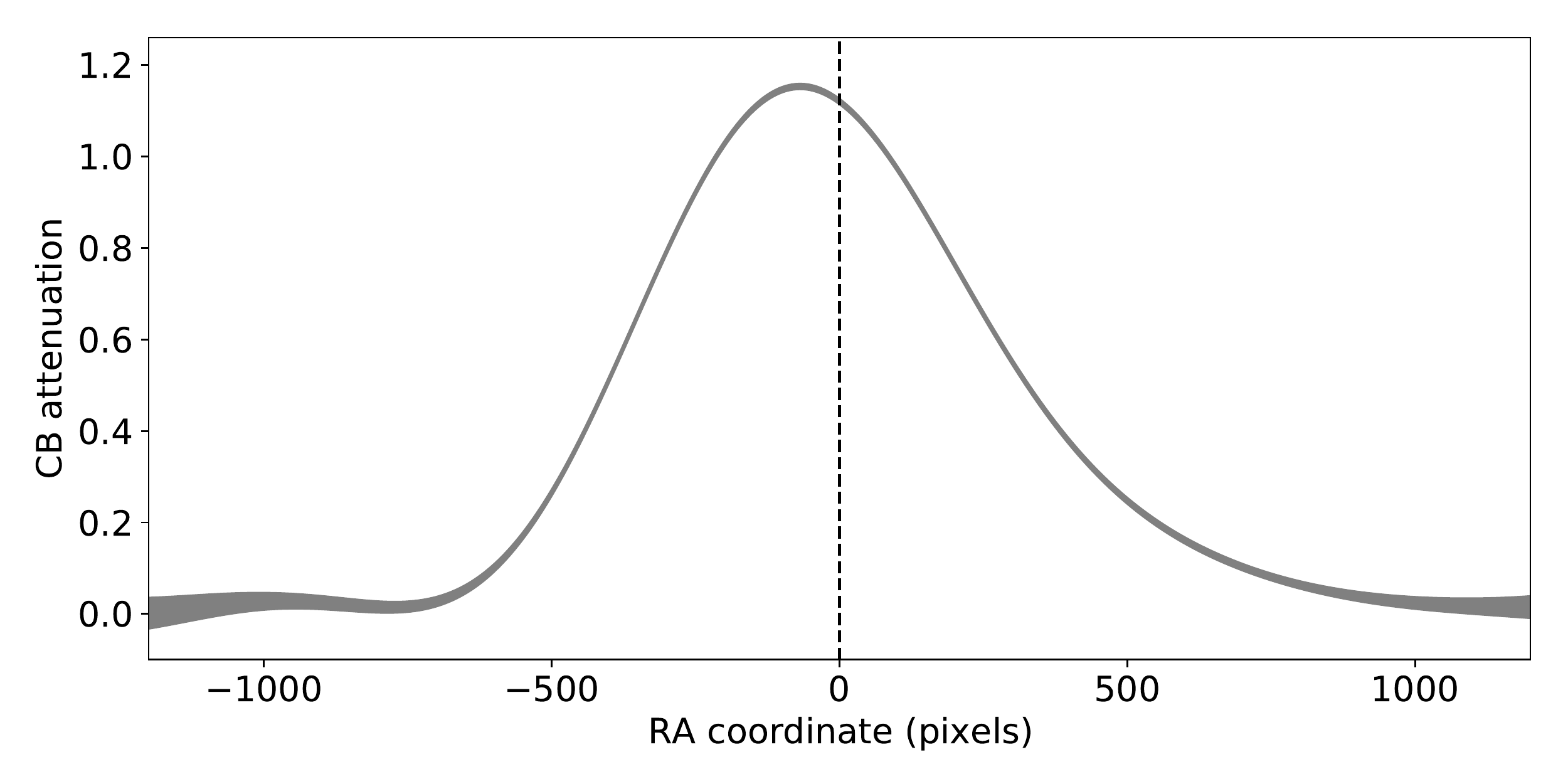}~\includegraphics[width=\columnwidth]{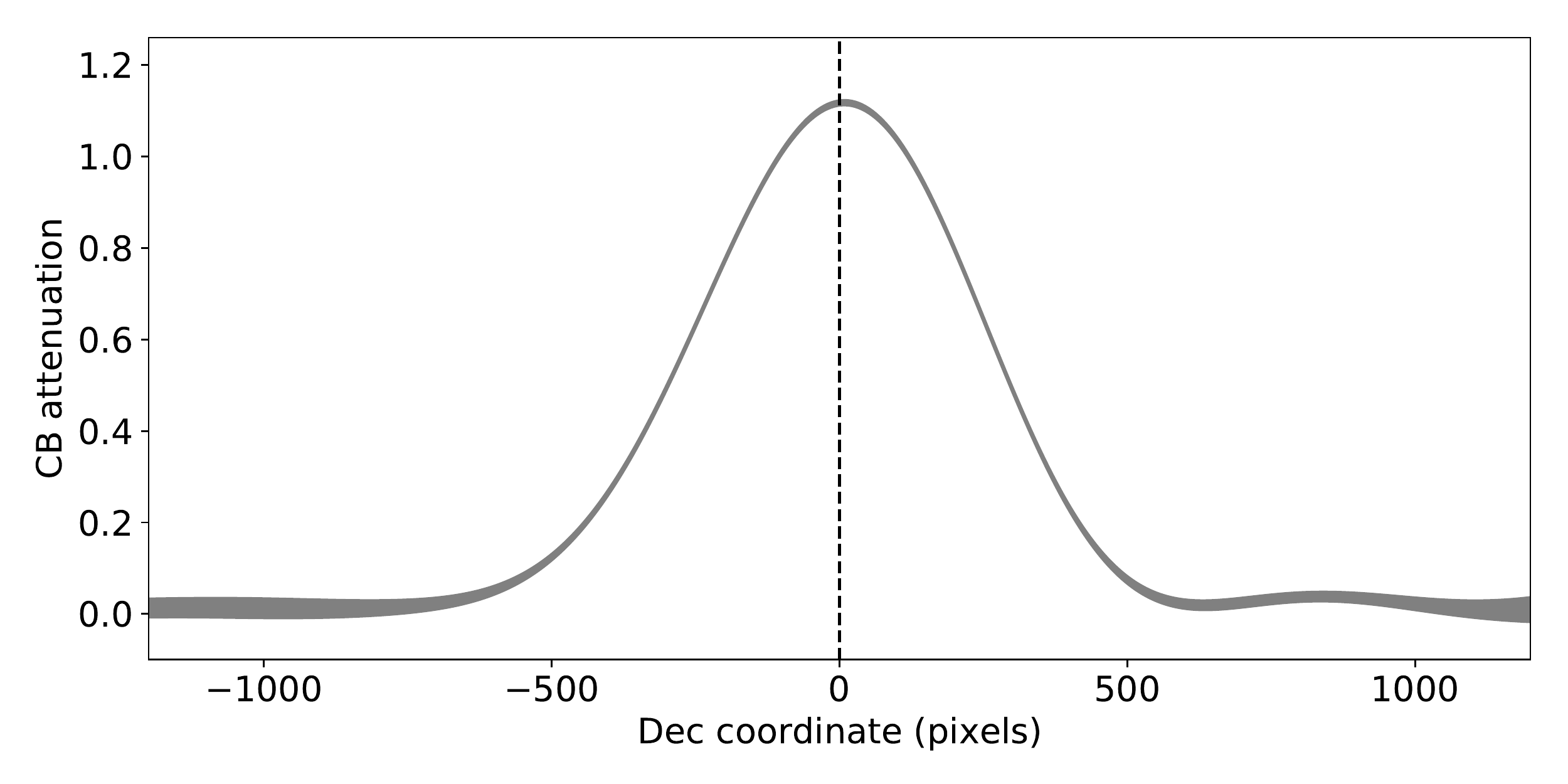}
    \caption{Implementation of Gaussian process regression for compound beam 01. Top left: relation of Apertif/NVSS integrated flux density for all the sources observed in the compound beam 01 (the size of the markers is proportional to the value). Top right: the GP regression built on these data. Bottom row: slices of the predicted CB shape ($\pm1\sigma$) along RA and Dec coordinates.}
    \label{fig:gp}
\end{center}
\end{figure*}

In Figure~\ref{fig:gp} the implementation of Gaussian process regression for compound beam 01 is shown. This beam is located at one of the  corners of the Apertif field of view and, therefore, is expected to have a distorted shape (e.g., Fig\,25 in~\cite{2022A&A...658A.146V}). The top left panel shows the scatter plot of the relation $\epsilon$ for all the sources cross-matched in the CB\,01 -- 7154 measurements taking values from 0.0004 to 2.6 (the sizes of the markers in the plot are proportional to the value). These data become a training sample for GP regression, of which the result is plotted in the top right panel. The asymmetry of the CB shape can be clearly seen from the figure. 
The bottom panels show slices of the GP surface along RA and Dec coordinates, where also the displacement of the peak with respect to the pointing center is seen. Both the shape distortion and the shift of the maximum are important to account for during a primary beam correction procedure. Moreover, it was found that the CB shapes change in time, which might be related to the re-adjustment of the signals from focal Vivaldi antennas (beam weights tuning procedure) as well as to other system drifts. This effect will be investigated for future data releases, while in this work we use the averaged profiles of the CBs based on the full year observational data. 
Since there are many measurements of $\epsilon$ for a given CB, this makes the GP training process very slow. To speed it up we used the k-means clustering of these measurements into 1000 clusters. We additionally checked that GP regression obtained on such a clustered sample does not differ significantly from the one obtained on a full sample.

% However, the CB shapes change in time, so we construct a GP regression for each observation. To get enough data for GP training (we require at least 500 measurements of $\epsilon$) we take the nearby observations within one week of the current one. This choice is based on the fact  that within a two weeks time spat the beamweights usually remain unchanged. This interval is being iteratively increased if there are still less than 500 measurements. 

The proposed approach has multiple advantages. First, this method provides a ready-to-use correction image for a given CB image in the sense that it does not require  knowledge of a frequency or antenna dependence of a CB shape. Next, the approach is free from approximation errors of analytical models. Indeed, an evolution of a CB slope or a shift of the peak of a CB with respect to image center may be hard to model analytically. 
Third, since a CB shape changes in time, one can select the data within a limited time period, and, therefore, obtain a CB correction for a given observation period. Finally, all possible sources of  bias contributing to an individual CB image turn into just one - the one of the NVSS catalog itself, which is assumed to be negligible throughout this work. 

For the ADR1, there are two versions of these beams released: the "original" primary beam images which provide a correction that matches the NVSS flux scale by design, and the "normalized" primary beam images are normalized to a peak response of one.
The data release documentation contains information on how
to best apply these primary beam images to the Apertif data, including how to scale the primary beam images with frequency.
The accuracy of the flux scale with the GPR beams is discussed in Section
\ref{sec:flux_scale}. The source code used to produce the CB models is available on GitHub\footnote{\url{https://github.com/akutkin/abeams}}.

%%%%%%%%%%%%%%%%%%%%%%%%%%%%%%%%%%%%%%%%%%%%%%%%%%%%%%%%%%%%
\section{Mosaicing of the images}
\label{sec:mosaicing}
Here we describe the procedure for the creation of the corrected images taking advantage of mosaicing with neighboring beams in order to achieve nearly uniform noise across the field, important for the source extraction and the catalog.

As mentioned above and described by \cite{2022A&A...658A.146V} and Hess et al. (in prep.), the adopted layout of CBs locations is such that neighboring CBs of Apertif overlap. Thus, the sensitivity for each field can be increased by making a mosaic of these images. 
For every CB image we take those closest surrounding images with their centers separated less than 0.7 degrees from the current CB center. Since only the images released in ADR1 are used in this work, some CB images have less than six surrounding ones. Moreover, some appear to be complete orphans -- there are 38 CB images without any neighbors. This circumstance leads to a patchy sky coverage and an increase of noise at the edges of such orphan CB image. The number of good quality CB images will significantly increase in future Apertif data releases, in particular due to a new pipeline capable of performing direction-dependent calibration. 
After selecting a central CB image and its neighbor beams, the smallest Gaussian restoring beam which encloses all individual restoring beams of the selected image is calculated. This is typically a few arcseconds larger than the individual ones. All the images are then convolved to this common restoring beam. Next, each image is corrected for the corresponding CB shape clipped at the CB attenuation level of 0.1, which was chosen to avoid very high noise levels at  mosaic edges.
After this, the images are re-projected onto a common projection center and stacked together into a mosaic image. Every mosaic image therefore covers up to 3.7 square degrees and represents a stack of individual CB images. The resulting mosaic images are used for source finding. We also generated a multi-order coverage map (MOC) of the survey, indicating the overall sky coverage of the mosaics of 970 square degrees. 
The source code for mosaicing is available on GitHub\footnote{\url{https://github.com/akutkin/amosaic}}. 

\begin{figure*}
    \centering
    \includegraphics[width=\linewidth]{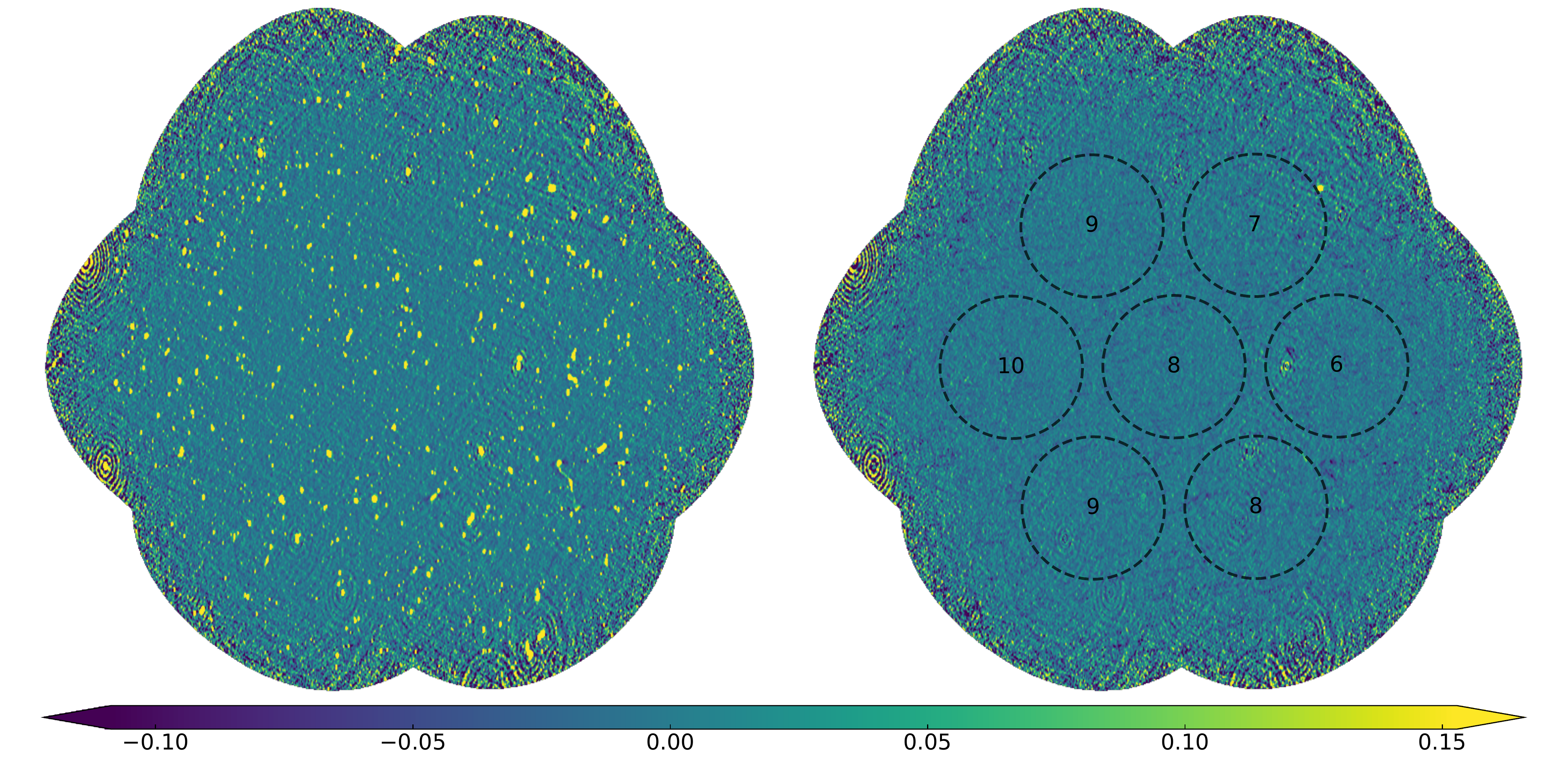}
    \caption{Mosaic example. Mosaic of 57 individual CB images with the center at RA=29.26, Dec=34.55 degrees (left panel) and the residual noise map (right panel). Seven dashed circles show the positions of compound beams, and the numbers inside indicate the quantity of the corresponding CB images used for mosaicing. The color bar scale is given in mJy/beam.}
    \label{fig:mosaic}
\end{figure*}

An example of the mosaic image made of 57 individual CB images of a region of the medium-deep survey (all released CB images of several observations for these coordinates) is shown in Figure~\ref{fig:mosaic}. On the right panel of the figure the residual map after source extraction  is shown (see the next section). The dashed circles and the numbers indicate the positions and the number of  CB images used for making the mosaic. The image has one of the lowest local noise levels in its central region reaching \ujybeam{17}. Imaging artifacts around some brighter sources clearly visible in the residual image on the right are caused by direction-dependent effects, which will be solved for in upcoming data releases with an improved pipeline.

%%%%%%%%%%%%%%%%%%%%%%%%%%%%%%%%%%%%%%%%%%%%%%%%%%%%%%%%%%%%

\section{Source extraction and catalog}
\label{sec:cataloging}

In each mosaic, the source extraction is done using the Python Blob Detector and Source Finder \citep[PyBDSF;][]{2015ascl.soft02007M}, which is known to be one of the best performing among continuum source finding software~\citep{hopkins_ASKAP_source_finders}. It produces a source catalog from an image by grouping together emission components modeled by Gaussians. Such a group is represented by an ``island'', a region of pixels having flux above a certain threshold. With this approach PyBDSF efficiently groups emission blobs into sources. We set the default peak-to-noise source detection threshold to 5 and the island boundary threshold to 3. The background noise level was determined inside a sliding box of 128 pixels across the image with a step of 16 pixels (\texttt{rms\_box}=(128,16)). The box size was adjusted dynamically, decreasing to 16 pixels in the regions of high background noise (\texttt{adaptive\_rms\_box}=\textit{True}, \texttt{rms\_box\_bright}=(16,4)). 
The imaging artifacts near bright sources occur mostly due to direction-dependent effects and appear as concentric circles around a source, affecting the estimated noise level. The size of these rings ranges from a few to a few tens of the size of the restoring beam. These characteristic scales were used for the size of the sliding box to properly detect the background noise variations. We tested a few other combinations of the \texttt{rms\_box} and \texttt{rms\_box\_bright} parameters and used one resulting in a lower number of false detections.
We note, that a background RMS estimate within a small box might experience additional variations due to confusion, however this effect is small for Apertif which has a confusion limit of $\sim5$\ujybeam{}.
A more justified and automated approach for the parameters setup should be developed in future.

Thus, for every mosaic image we obtain a source list which includes coordinates, size, signal-to-noise ratio and other fitting parameters. We used the source position and size from the PyBDSF output to identify duplicate sources in different mosaic images. Finally, the source lists for each mosaic image were combined together and duplicate sources having lower signal-to-noise were removed. 

%%%%%%%%%%%%%%%%%%%%%%%%%%%%%%%%%%%%%%%%%%%%%%%%%%%%%%%%%%%%

\subsection{Extra filtering}

In Apertif images a ``ghost'' source is present at the image center. This is the result of some wide band radio frequency interference (RFI) within the backend. This RFI signal has a random phase which averages to 0 and produces a fake source during imaging. Noting this we removed those ghost sources from the catalog by matching their coordinates with the CB centers.

We also note that in spite of the strong validation criteria applied for ADR1 and the carefully adjusted PyBDSF parameters some artifacts were still detected as extended sources. They tend to have an order of magnitude higher coordinates uncertainties and thus can be easily identified. We inspected visually the sources with either Right Ascension or Declination errors higher than \asec{10} and dropped the detected artifacts. 

%%%%%%%%%%%%%%%%%%%%%%%%%%%%%%%%%%%%%%%%%%%%%%%%%%%%%%%%%%%%
\subsection{Catalog format}
\label{sec:cat_format}
The resulting catalog contains \nsources sources. Almost 90\% of the sources are modeled with a single Gaussian ('S'-flag), and around 10\% are classified as extended ('M'). An example of the catalog structure is shown in Table~\ref{tab:catalog}. The columns designations are: 
\\(1): Apertif source name; 
\\(2,4): RA and Dec; 
\\(3,5): RA and Dec errors (see Appendix~\ref{sec:errors}); 
\\(6,8): Total and Peak flux density; 
\\(7,9): Integrated and Peak flux density uncertainties, calculated as described in Appendix~\ref{sec:errors};
\\(10,12,14): Deconvolved major/minor source size and position angle. A source size value 0.0 means that the PSF can not be deconvolved from the fitted source along the given axis, and the corresponding uncertainty, represents an upper size limit estimate. The position angle is given in degrees from -90 to 90 measured relative to the north celestial pole, turning positive into the direction of the right ascension. When both major and minor size are 0.0 the position angle is omitted; 
\\(11,13,15): Uncertainties of major/minor source size and position angle calculated as described in Appendix~\ref{sec:errors}.;
\\(16): local background noise RMS; 
%\\(17): Mosaic name in TaskID\_beam format; 
\\(17): Source type as classified by PyBDSF ('S' -- an isolated source fitted with a single Gaussian; 'C' -- sources that were fit by a single Gaussian but are within an island of emission that also contains other sources, and 'M' -- sources fitted with multiple Gaussians).

%%%%%%%%%%%%%%%%%%%%%%%%%%% CATALOG TABLE -- horizontal %%%%%%%%
% \setlength{\tabcolsep}{3pt}
% %--- table ---
% \ctable[topcap,center,sideways,
% doinside=\tiny,%\small,%
% caption = {A sample of entries from the catalog. Descriptions of the columns are given in Section~\ref{sec:mosaicing}. The full table with \nsources ~entries will be available in electronic form through CDS.},
% label = tab:catalog]{c c c c c c c c c c c c c c c c c c c}{}
% {
%  \FL 
% Name & RA & $\sigma_\mathrm{RA}$ & Dec & $\sigma_\mathrm{Dec}$ & $S_\mathrm{total}$ &  $\sigma_{S_\mathrm{total}}$ &  $S_\mathrm{peak}$ &  $\sigma_{S_\mathrm{peak}}$ &  Maj &  $\sigma_\mathrm{Maj}$ & Min &  $\sigma_\mathrm{Min}$ & PA & $\sigma_\mathrm{PA}$ & RMS & Mosaic & S\_Code
%  \NN 
% {\changeone{(APTF)}} & [\adeg{}] & [\asec{}] & [\adeg{}] & [\asec{}] & [mJy] & [mJy] & [\mjybeam{}] & [\mjybeam{}] & [\asec{}] & [\asec{}] & [\asec{}] & [\asec{}] & [\adeg{}] & [\adeg{}] & [\mjybeam{}] &  & {\changeone{(S/M/C)}} & {}
%  \NN 
% (1) & (2) & (3) & (4) & (5) & (6) & (7) & (8) & (9) & (10) & (11) & (12) & (13) & (14) & (15) & (16) & (17) & (18)
%  \ML
% \input{tab_cat_rev}
% \LL
% }
%%%%%%%%%%%%%%%%%%%%%%%%%%%%%%%%
\begin{table*}
    {\scriptsize
    \caption{Continuum source catalog. \label{tab:catalog}}
    \begin{tabular}{lcccccccccccccccc}
    \toprule
    Name & RA & $\sigma_\mathrm{RA}$ & Dec & $\sigma_\mathrm{Dec}$ & $S_\mathrm{total}$ &  $\sigma_{S_\mathrm{total}}$ &  $S_\mathrm{peak}$ &  $\sigma_{S_\mathrm{peak}}$ &  Maj &  $\sigma_\mathrm{Maj}$ & Min &  $\sigma_\mathrm{Min}$ & PA & $\sigma_\mathrm{PA}$ & RMS & S\_Code \\
    ~  & [\adeg{}] & [\asec{}] & [\adeg{}] & [\asec{}] & mJy & mJy & mJy/bm & mJy/bm & [\asec{}] & [\asec{}] & [\asec{}] & [\asec{}] & [\adeg{}] & [\adeg{}] & $\muup$Jy/bm &  (S/M/C) \\
    (1) & (2) & (3) & (4) & (5) & (6) & (7) & (8) & (9) & (10) & (11) & (12) & (13) & (14) & (15) & (16) & (17) \\
     \midrule
     APTF\_J013717+362720 & 24.3249 & 2.0 & 36.4558 & 3.0 & 0.26 & 0.07 & 0.18 & 0.03 & 11.6 & 11.6 &  8.2 &  5.9 &  44.9 &  70.4 &  28.3 &  S \\
 APTF\_J013718+302840 & 24.3256 & 1.0 & 30.4779 & 1.0 & 5.49 & 0.36 & 4.46 & 0.25 &  0.0 &  9.4 &  0.0 &  6.3 &   0.0 & \dots & 117.1 &  S \\
 APTF\_J013718+364631 & 24.3276 & 1.1 & 36.7756 & 1.2 & 0.77 & 0.07 & 0.68 & 0.04 &  0.0 & 11.6 &  0.0 &  6.5 &   0.0 & \dots &  29.6 &  S \\
 APTF\_J013718+362056 & 24.3280 & 1.3 & 36.3491 & 1.5 & 0.54 & 0.07 & 0.50 & 0.04 &  0.0 & 11.9 &  0.0 &  5.3 &   0.0 & \dots &  32.4 &  S \\
 APTF\_J013718+355643 & 24.3286 & 1.9 & 35.9453 & 3.2 & 0.31 & 0.08 & 0.23 & 0.04 & 13.9 & 11.4 &  8.1 &  0.7 &  -6.9 &  28.0 &  34.6 &  S \\
 APTF\_J013718+364753 & 24.3289 & 1.3 & 36.7982 & 1.7 & 0.52 & 0.06 & 0.44 & 0.03 & 13.9 &  9.0 &  0.0 &  5.2 &  21.2 &  37.6 &  30.1 &  S \\
 APTF\_J013718+344346 & 24.3289 & 2.0 & 34.7297 & 3.5 & 0.30 & 0.08 & 0.23 & 0.04 & 16.0 & 11.8 &  0.0 &  9.3 & -29.4 &  34.7 &  37.2 &  S \\
 APTF\_J013718+361706 & 24.3289 & 1.2 & 36.2851 & 1.0 & 1.38 & 0.12 & 0.99 & 0.06 &  0.0 & 14.8 &  0.0 &  3.0 &   0.0 & \dots &  32.9 &  M \\
 APTF\_J013719+350259 & 24.3301 & 1.0 & 35.0499 & 0.9 & 1.20 & 0.08 & 1.06 & 0.06 &  6.0 &  2.5 &  2.8 &  3.2 & -79.7 &  80.1 &  26.4 &  S \\
 APTF\_J013719+354156 & 24.3303 & 1.3 & 35.6991 & 1.6 & 0.47 & 0.06 & 0.43 & 0.04 &  4.8 &  8.7 &  4.3 &  0.4 & -28.5 &  50.4 &  30.7 &  S \\
 APTF\_J013719+370339 & 24.3309 & 1.5 & 37.0609 & 1.9 & 0.54 & 0.09 & 0.50 & 0.05 &  6.1 &  9.0 &  0.0 &  5.2 &  48.1 &  63.9 &  46.8 &  S \\
 APTF\_J013719+353747 & 24.3311 & 4.3 & 35.6298 & 2.6 & 1.48 & 0.16 & 0.65 & 0.04 & 60.8 & 12.8 &  0.0 &  3.6 & -33.4 &   8.6 &  30.6 &  M \\
 APTF\_J013719+302519 & 24.3312 & 1.5 & 30.4221 & 2.1 & 1.57 & 0.26 & 1.33 & 0.13 & 10.1 &  8.6 &  0.0 &  7.9 &  49.7 &  52.8 & 125.5 &  S \\
 APTF\_J013719+372027 & 24.3317 & 2.1 & 37.3411 & 3.0 & 0.27 & 0.08 & 0.22 & 0.04 &  0.0 & 21.7 &  0.0 & 11.1 &   0.0 & \dots &  36.8 &  S \\
 APTF\_J013719+345326 & 24.3321 & 2.3 & 34.8907 & 3.8 & 0.20 & 0.06 & 0.15 & 0.03 & 14.0 & 13.3 &  3.7 &  7.6 & -39.4 &  54.2 &  27.5 &  S \\
 APTF\_J013719+371841 & 24.3321 & 2.0 & 37.3116 & 3.2 & 0.41 & 0.10 & 0.25 & 0.04 & 16.6 & 10.7 & 11.2 &  5.0 & -13.7 &  36.6 &  38.3 &  S \\
 APTF\_J013720+364236 & 24.3337 & 1.4 & 36.7100 & 1.8 & 0.36 & 0.05 & 0.31 & 0.03 &  5.9 &  8.7 &  4.5 &  1.0 & -80.7 &  77.8 &  26.0 &  S \\
 APTF\_J013720+310549 & 24.3338 & 2.2 & 31.0971 & 4.5 & 0.86 & 0.22 & 0.52 & 0.08 & 28.8 & 17.0 &  0.0 &  8.6 & -20.9 &  31.5 &  90.4 &  S \\
 APTF\_J013720+363404 & 24.3345 & 1.7 & 36.5679 & 2.4 & 0.22 & 0.05 & 0.21 & 0.03 &  0.0 & 15.2 &  0.0 &  2.4 &   0.0 & \dots &  26.0 &  S \\
 APTF\_J013720+373637 & 24.3347 & 1.1 & 37.6103 & 1.1 & 1.85 & 0.15 & 1.30 & 0.08 &  0.0 & 14.9 &  0.0 &  7.4 &   0.0 & \dots &  50.9 &  S \\
\bottomrule

    \end{tabular}
    \tablefoot{Sample of the catalog records. Descriptions of the columns are given in Section~\ref{sec:cat_format}. The full table containing \nsources entries will be available in machine-readable format through CDS.}
    }
\end{table*}

\subsection{Correction for spectral index}\label{sec:si_correction}

The central frequency of the NVSS survey observing band differs from the Apertif one by 45 MHz. Within the aforementioned approach to the primary beam correction,  source fluxes would undergo a systematic offset depending on their spectral index\footnote{In this paper, the spectral index $\alpha$ is defined through $S \propto \nu^{{\alpha}}$, where $S$ is flux density and $\nu$ the frequency.} (see also Section~\ref{sec:SI}). 
A flux correction for a source with a spectral index of --0.7 will be $\xi(-0.7) = 1.023$, or 2.3 percent (the dependence of the correction factor on the spectral index is weak: $\xi(-0.5) = 1.016$, $\xi(-1) = 1.033$). In order to eliminate this bias we apply the correction factor 1.023 to all sources in the catalog. 

\subsection{Noise properties}
Analysis of the noise properties is important for the discussion that will follow and to quantify the completeness of the catalog. 
The background noise varies strongly from one mosaic image to another as well as within a single mosaic image.
This can be seen in the residual image on the right panel of Figure~\ref{fig:mosaic}. The noise in this image reaches \ujybeam{17} in the central region and goes to above \ujybeam{100} at the edges.
For the catalog, we effectively select sources from a central part of every individual mosaic because the sources at the edges are typically duplicated/substituted by ones from central region of another adjacent mosaic having higher signal-to-noise. This means, that a more informative parameter is the noise level around the cataloged sources, not in an entire mosaic map. The relevant noise measure is the median background RMS of the sources provided by PyBDSF (column 16 in the catalog) which is \ujybeam{44}. We also note that because of the ADR1 validation criteria, many CB images have fewer than six neighbors, which implies a higher noise level at part of the edges. 
%%%%%%%%%%%%%%%%%%%%%%%%%%%%%%%%%%%%%%%%%%%%%%%%%%%%%%%%%%%%

\subsection{Reliability and completeness}
\label{sec:completeness}

\begin{figure}
    \centering
    \includegraphics[width=\columnwidth]{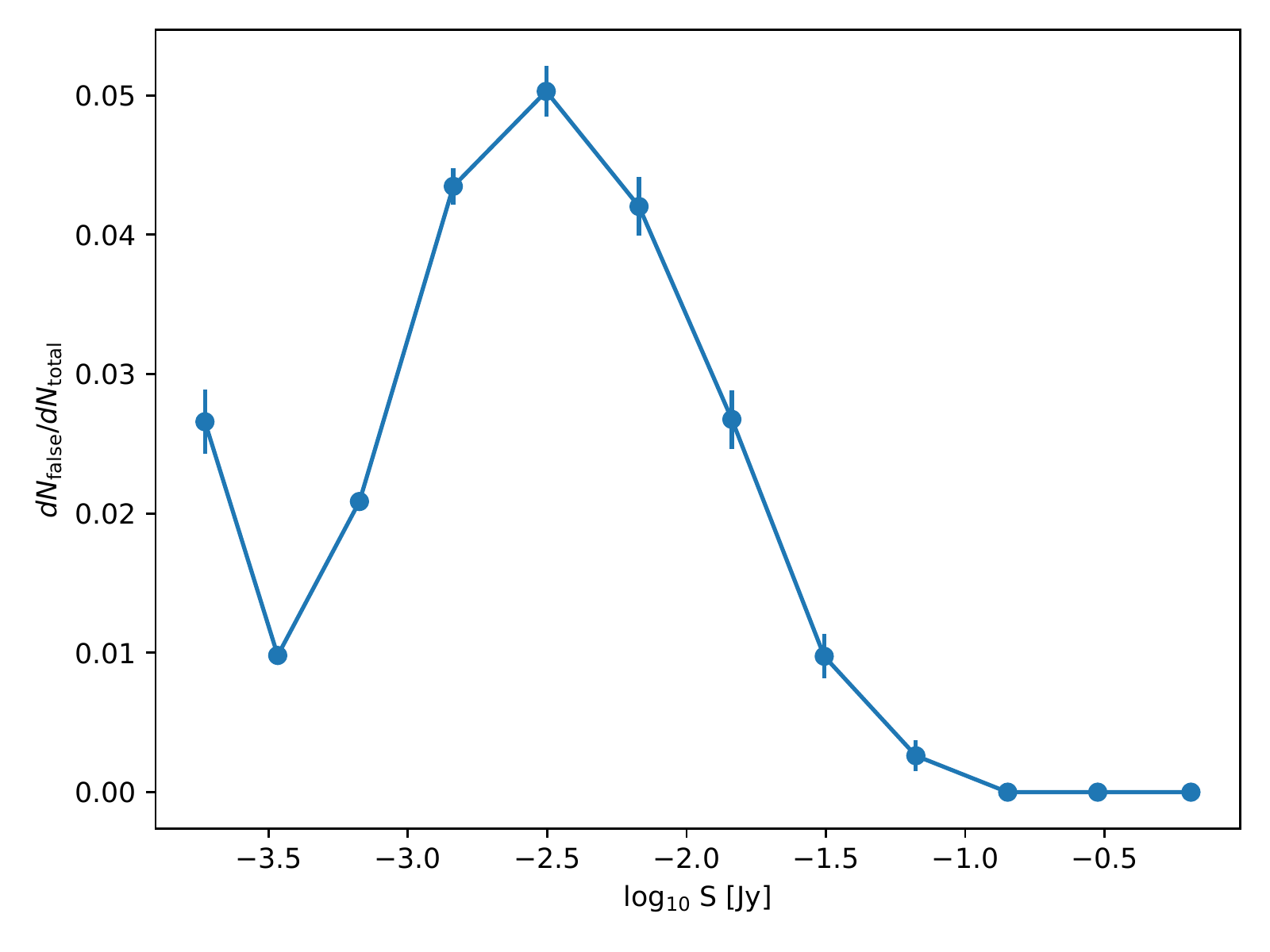}
    \caption{Relative number of false detections as a function of flux density.}
    \label{fig:false_positives}
\end{figure}

To estimate the reliability of the catalog through the number of false  detections, we run the same source finding procedure on the inverted (multiplied by --1) mosaic images. Within the assumption of a symmetric noise distribution this gives a rough estimate of the number of fake sources in the original images. 
We compiled the ``inverted'' catalog in the same way as the original one. It contains 6108 sources or 2.4\% of the number of sources in the main catalog. In Figure~\ref{fig:false_positives} the false detection rate (FDR) is plotted against flux density where the FDR is the relative number of sources from inverted catalog to the number of the sources from the original catalog taken in various flux density intervals. The peak of the FDR around a few mJy is likely related to the typical flux of imaging artifacts around bright sources, while artifacts around fainter sources are below the catalog detection limit. We also found that 64\% of the false detections reside within \amin{2} of sources from the original catalog that are brighter than \mjy{50}. Overall, among the sources weaker than \mjy{30}, a few percent may be false detections, while brighter sources are even more reliable detections.

\begin{figure}
    \centering
    \includegraphics[width=\columnwidth]{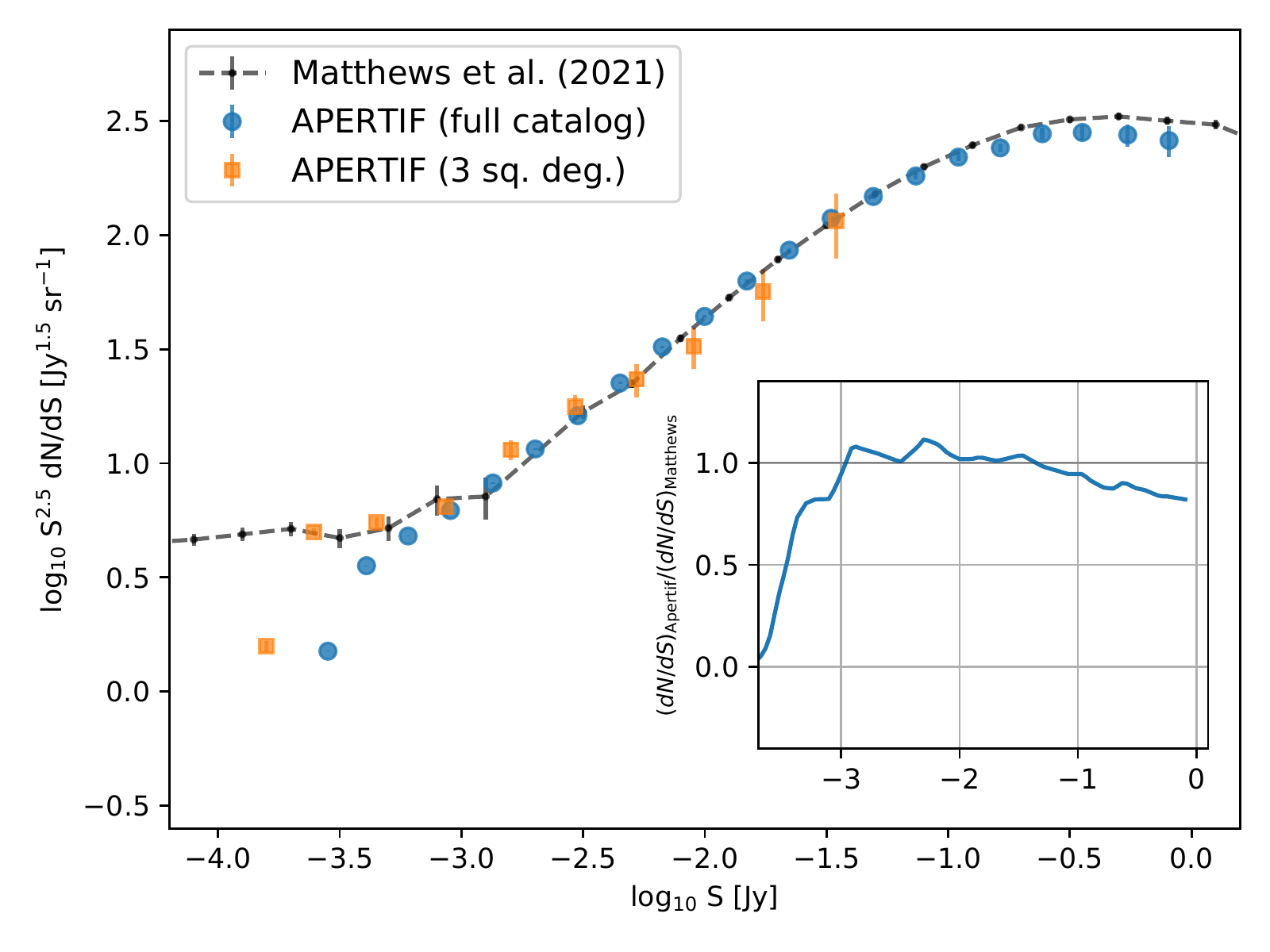}
    \caption{Differential source counts. The source counts estimates for the full catalog area (circles) and for the low-noise mosaic image (Figure~\ref{fig:mosaic}, squares) plotted along with the measurements of~\citet{2021ApJ...909..193M} (dot-dashed curve). The ratio $(dN/dS)_{\rm Apertif} / (dN/dS)_{\rm Matthews}$ is shown in the inset plot.}
    \label{fig:source_counts}
\end{figure}

We estimate the completeness of the catalog by comparing the differential source counts corrected for the false positive detection rate to the one by \citet{2021ApJ...909..193M}. The comparison is shown in Figure~\ref{fig:source_counts}. The ratio $(dN/dS)_{\rm Apertif} / (dN/dS)_{\rm Matthews}$ is shown in the inset plot.
Slightly lower counts of bright sources in the Apertif catalog are dictated by the ADR1 validation criteria, namely by the lack of the released images with stronger sources due to them having imaging artifacts (see Section~\ref{sec:data}). Another effect might be a flux underestimating for the bright sources due to some of them being partially resolved with Apertif.
On the weaker end, we find that at the 1 mJy level the catalog completeness is about 96\%, dropping to $\sim$75\% at 400\,$\muup$Jy level. 
We underline that this level is calculated for the full catalog. In reality the number of weak sources and hence \textit{local} completeness in a given sky direction depends on the local noise level in an image. To illustrate this, in Figure~\ref{fig:source_counts} we plot the source counts calculated for the low-noise mosaic image shown in Figure~\ref{fig:mosaic} which covers approximately 3 square degrees. The corresponding ``local'' catalog is then complete down to the \mjy{0.1} level.

\subsection{Accuracy of the flux scale}
\label{sec:flux_scale}

In order to assess the flux scale accuracy we cross-matched the Apertif catalog with the NVSS one using a \asec{20} matching radius. This results in 44524 common sources, for which the integrated flux density differs by less than 1\% on average with median ratio $S_{int,APTF}/S_{NVSS} = 1.027$. Note that the applied correction factor for the spectral index is 1.023, as described in Section~\ref{sec:si_correction}. Obviously, such a good agreement is a consequence of the applied method for the primary beam correction (section~\ref{sec:cbeams}). This comparison serves as a firm additional check for the flux scale of the Apertif catalog as a whole, especially taken that it has a different angular resolution than the NVSS catalog. The standard deviation of the above relation is 0.345 indicating a significant scatter. This spread must be related to the variability of the sources, different angular resolution and sensitivity of the catalogs, and errors of the flux measurements. 

\begin{figure}
    \centering
    \includegraphics[width=\linewidth]{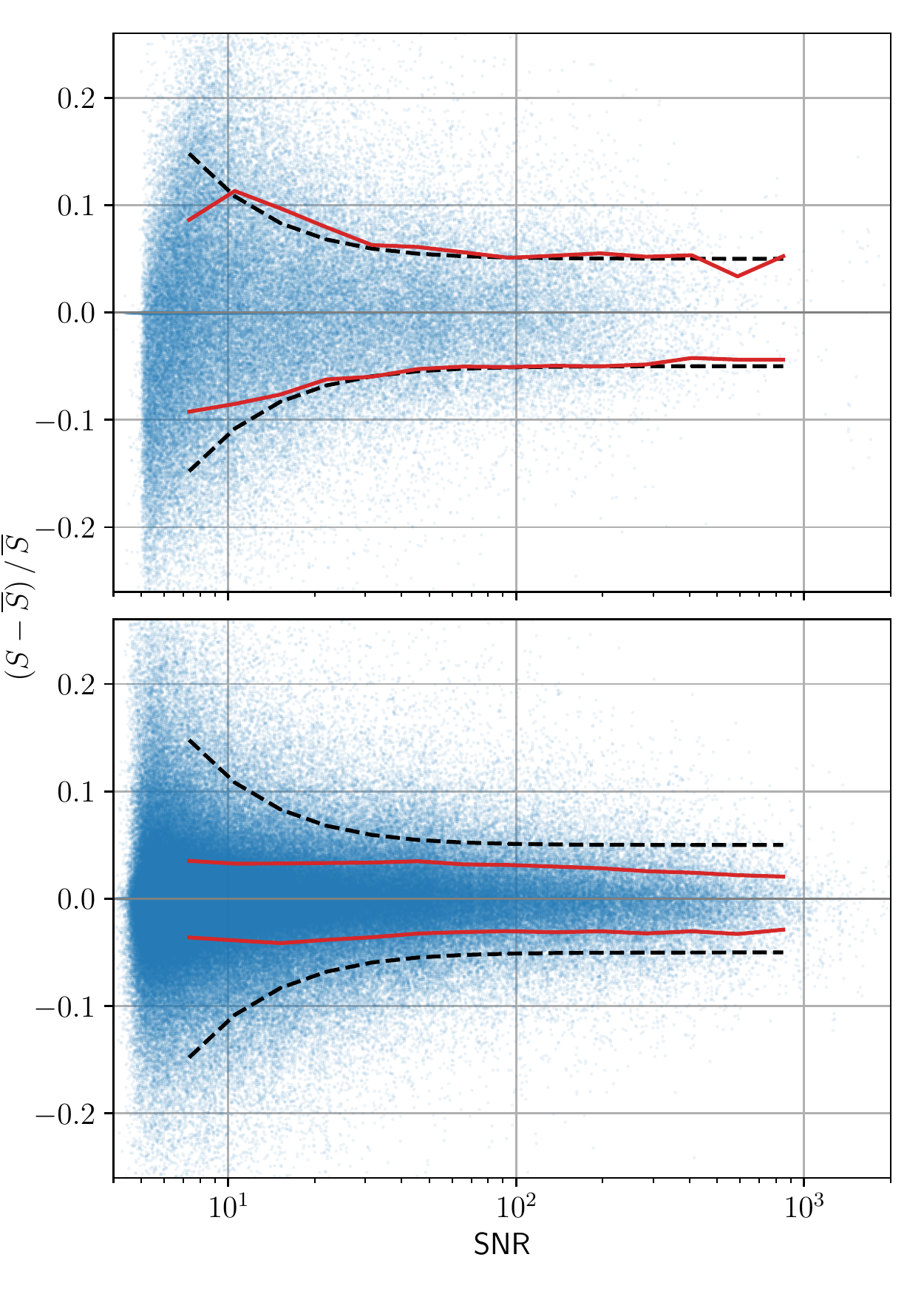}
    \caption{Flux measurements precision. Relative difference of the peak flux measurements over time (top) and between mosaic images (bottom) against signal-to-noise ratio of the sources. The dashed curves indicate average relative peak flux error. The solid curves show 16 and 84 percentiles.}
    \label{fig:flux_acc}
\end{figure}

In order to estimate the precision of the flux measurements we compare these measurements obtained from the CB images of the medium-deep fields which have at least 5 pointings. The variability of flux measurements within individual Apertif images is shown on the top panel of Figure~\ref{fig:flux_acc}, where the relative difference of the peak flux and the mean peak flux is plotted against signal-to-noise ratio (SNR) of the sources. We split the full SNR range into 15 bins, then for every bin calculated the average relative peak flux error and 16/84 percentiles of the distribution showed with the dashed and solid curves respectively. 
We conclude that the flux measurements precision is better than 10\% consistent with the expected errors coming from data cross-calibration. 

Finally, we check the intrinsic flux scale consistency by comparing the fluxes of the sources from different overlapping mosaics. When a source was detected in multiple images, only the corresponding record with highest signal-to-noise was stored in the catalog. Thereby, comparing fluxes of those sources provide a direct probe of the applied primary beam correction and overall correctness of the mosaicing procedure (Sections~\ref{sec:cbeams}~and~\ref{sec:mosaicing}).
For these sources we again consider the relative difference of the peak flux and its mean value. The distribution is plotted in the bottom panel of Figure~\ref{fig:flux_acc} against SNR of the sources. The scatter indicates that the error during primary beam correction and mosaicing does not exceed 5\%. 
We take this into account for the flux error values reported in Table~1 as described in Appendix~\ref{sec:errors}

%%%%%%%%%%%%%%%%%%%%%%%%%%%%%%%%%%%%%%%%%%%%%%%%%%%%%%%%%%%%
\subsection{Astrometry} 
\label{sec:astrometry}

\begin{figure}
    \centering
    \includegraphics[width=\linewidth]{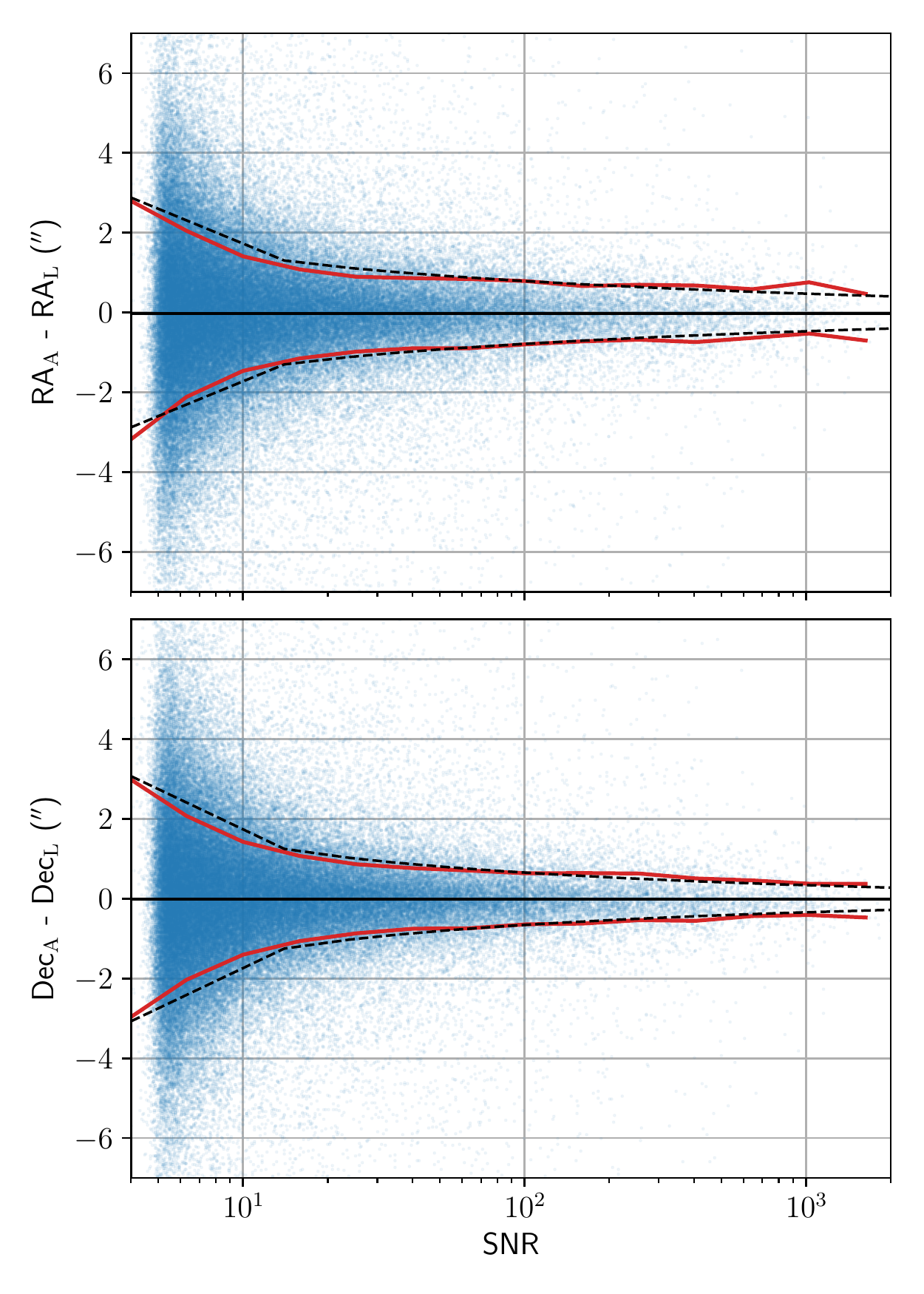}
    \caption{Astrometric accuracy. Coordinate offsets of the compact sources based on Apertif -- LOFAR cross-match (\textit{top}: Right Ascension, \textit{bottom}: Declination). The horizontal lines indicate median value, the solid curves show 16 and 84 percentiles, and the dashed curves show the corresponding hyperbola fit (see details in Appendix~\ref{sec:errors}).}
    \label{fig:aplo_coords}
\end{figure}

To estimate astrometric accuracy of the sky coordinates it is desirable to have a reference catalog with a comparable or higher angular resolution. And fortunately, there is one -- the first Apertif data release has a common sky area of 700 square degrees with the second LOFAR 150\,MHz data release~\citep{2022A&A...659A...1S}.
We cross-match the Apertif catalog with the LoTSS\,DR2 catalog. Note that the LoTSS images have about two times better angular resolution (\asec{6}) than those from Apertif, and some sources resolved by LOFAR might remain unresolved by Apertif. We used an intermediate matching distance of \asec{10} resulting in \aplo common sources and consider the coordinate offsets between Apertif and LoTSS\,DR2 sources. In Figure~\ref{fig:aplo_coords} the offsets distributions for Right Ascension and Declination are plotted against SNR of the sources. The median values are \asec{0.0} and \asec{0.0} respectively indicating absence of any systematic shifts between the catalogs. The 16/84 percentiles are shown with the solid curves justifying the chosen matching radius. The coordinate uncertainties for the catalog were calculated as described in Appendix~\ref{sec:errors}.

%%%%%%%%%%%%%%%%%%%%%%%%%%%%%%%%%%%%%%%%%%%%%%%%%%%%%%%%%%%%
\section{Scientific applications}
\label{sec:science}

In this section we discuss some scientific applications of cross-matching the Apertif catalog with the NVSS and LOFAR/LoTSS catalogs.  
Cross-matching of catalogs obtained at similar frequency but at distinct epochs allows one to identify transient sources which changed their flux density significantly between the epochs. Finding common sources between two contemporaneous catalogs at different frequencies provides information about their spectral behavior. 

\subsection{Long term transients}
\label{sec:transients}

As an example for transient detection, we selected the sources which are present in the NVSS, but have disappeared in the Apertif images. First, we exclude the Apertif sources separated by less then \asec{50} from a mosaic edge, where the image quality is lower due to the noise. Then we cross-matched the remaining Apertif sources with the NVSS  and obtained a list of more than 4000 NVSS sources missing in our catalog. The great majority of these sources have a total flux density below \mjy{3} and represent NVSS false detections. We filtered these sources by requiring NVSS signal-to-noise to be higher than 10. Another large group consists of multi-component emission islands which were grouped by PyBDSF into sources whose location mismatch with the corresponding NVSS source. To filter these sources we inspected the Apertif images and required the signal-to-noise ratio at the position of each NVSS source to be less than 6. This ensures that there is no significant emission in Apertif images at the position of an NVSS source. Finally, we obtained a list of 23 NVSS sources having NVSS signal-to-noise higher than 10 and which were not detected in the Apertif data. The list is presented in Table~\ref{tab:aptf_nvss}. 

\begin{figure}
    \centering
    \includegraphics[width=\linewidth]{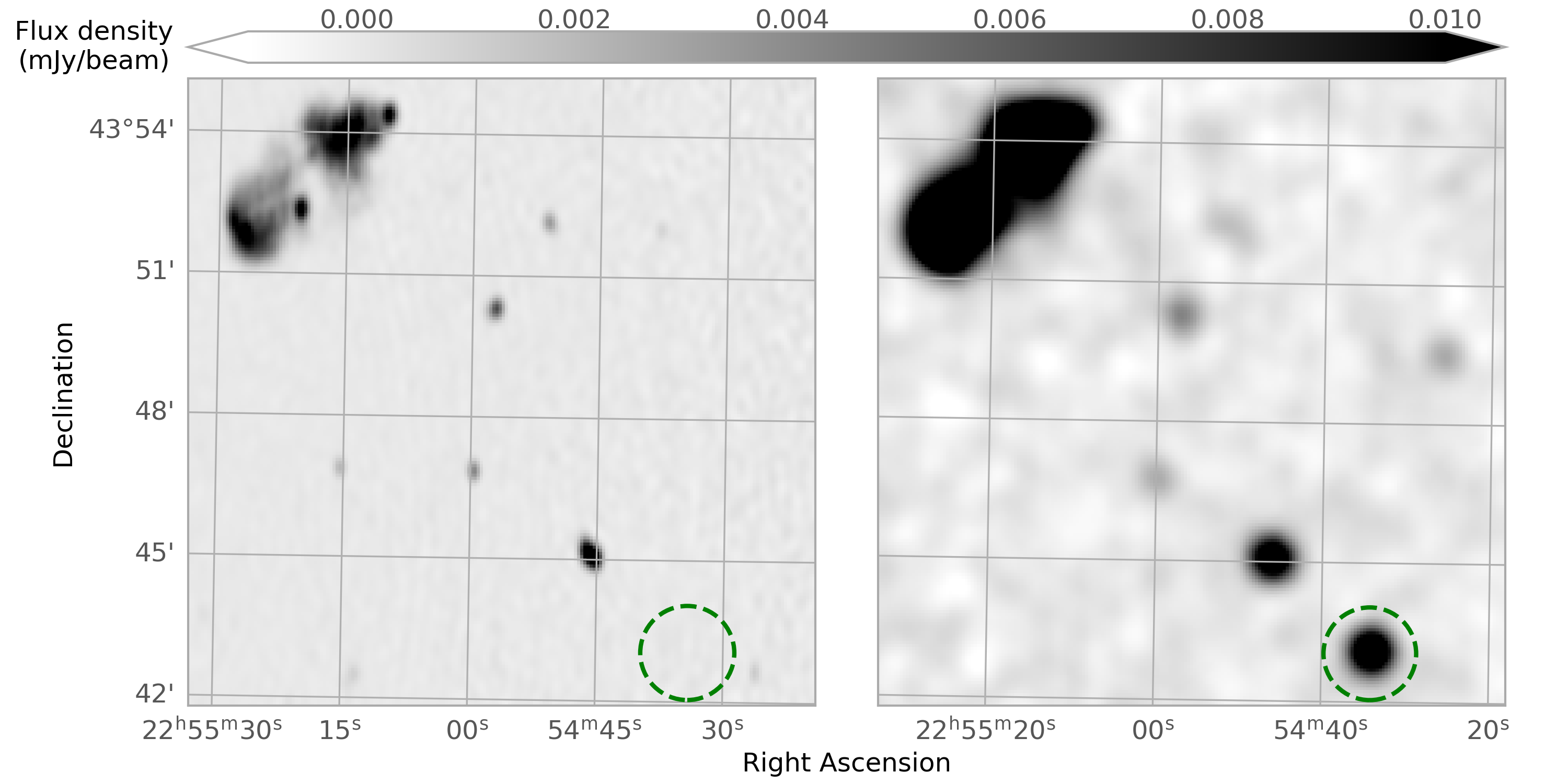}
    \caption{Apertif (left) and NVSS (right) images. The disappeared source is marked with the dashed circle and has total flux density of \mjy{$19.8\pm0.7$}.}
    \label{fig:aptf_nvss_images}
\end{figure}

Apparently, the source with the highest total flux density (NVSS:221119+385955) is a diffuse lobe of giant radio galaxy, resolved by Apertif. The next brightest source NVSS:224624+440950 is seen in the NVSS image (with a flux density of \mjy{19.8}) and is clearly absent in the Apertif one (Figure~\ref{fig:aptf_nvss_images}). The figure also nicely illustrates the differences in resolution and sensitivity  between the Apertif and NVSS surveys. The local noise  value in the Apertif mosaic at this position is \ujybeam{120}, implying that the source, if it is real and point-like, dropped its luminosity by a factor of~$\sim\,30$ (!). Such a change could, for example, have  happened if the source is a blazar whose jet changed its angle to the line of sight, causing a significant decrease of Doppler boosting. A factor of 30, however, is unusual even for extreme blazars~\citep[e.g.,][]{2018MNRAS.475.4994K}. Another possibility can be a supernova explosion. A detailed study of these objects (e.g., measuring redshifts, searching for counterparts in other bands, extra observations etc.) might be an interesting project for future research. We note that most sources in Table~\ref{tab:aptf_nvss} are relatively faint, therefore, some might be just variable sources picked up by NVSS when they were at maximum and missed by Apertif when they were at minimum. These transients, of course, deserve a separate study, while the aim of this section is only to demonstrate some usage examples for the new catalog. 

The above-mentioned procedure illustrates a use case example for the new catalog. We note, that a much larger sample of transient sources can be obtained by considering the significance of the flux density change and not just  sources that have disappeared. Such a detailed analysis, however, is beyond the scope of this paper. For the same reason we also do not consider the sources which are present in the Apertif survey and absent in the NVSS.

\begin{table}[]
    {\small
    \caption{NVSS sources not detected by Apertif. \label{tab:aptf_nvss}}
    \begin{tabular}{lcccc}
    \toprule
          NVSS &     RAJ2000 &    DEJ2000 &  S1\_4 &  e\_S1\_4 \\
               &   [\dg ]   &  [\dg ]   &  [mJy] &  [mJy] \\          
    \midrule    
         031640+405712 &   49.169000 &  40.953333 &   5.3 &     0.5 \\
 032228+414507 &   50.616792 &  41.752028 &   5.3 &     0.4 \\
 132422+493308 &  201.095583 &  49.552417 &  12.7 &     1.2 \\
 132919+300341 &  202.329542 &  30.061583 &   6.6 &     0.5 \\
 132929+302439 &  202.370833 &  30.410833 &   7.7 &     0.5 \\
 132932+303017 &  202.387417 &  30.504833 &   4.8 &     0.4 \\
 133426+305132 &  203.609125 &  30.859056 &   4.1 &     0.4 \\
 140300+520335 &  210.752917 &  52.059889 &   7.1 &     0.4 \\
 140305+520349 &  210.773917 &  52.063611 &   8.2 &     0.5 \\
 140840+511225 &  212.168875 &  51.207028 &   4.1 &     0.4 \\
 215545+374027 &  328.940750 &  37.674194 &   6.2 &     0.5 \\
 215548+374444 &  328.954083 &  37.745667 &  13.0 &     1.1 \\
 215549+373844 &  328.957000 &  37.645556 &   6.9 &     0.5 \\
 215722+381109 &  329.342625 &  38.185861 &   5.4 &     0.5 \\
 221119+385955 &  332.832208 &  38.998639 &  29.2 &     2.8 \\
 224624+440950 &  341.603625 &  44.164028 &  19.8 &     0.7 \\
 224642+394638 &  341.677208 &  39.777306 &   6.6 &     0.5 \\
 224655+442931 &  341.731083 &  44.492139 &  14.4 &     0.6 \\
 224701+440644 &  341.758042 &  44.112361 &   5.2 &     0.4 \\
 224804+441433 &  342.016750 &  44.242583 &   9.7 &     0.9 \\
 232359+303905 &  350.998458 &  30.651417 &  12.6 &     1.0 \\
 232430+304158 &  351.125000 &  30.699444 &   6.5 &     0.4 \\
 232434+304233 &  351.144542 &  30.709167 &   6.0 &     0.4 \\
    \bottomrule
    \end{tabular}
    \tablefoot{Columns: NVSS name, coordinates, total flux density and its error.}
    }
\end{table}

%%%%%%%%%%%%%%%%%%%%%%%%%%%%%%%%%%%%%%%%%%%%%%%%%%%%%%%%%%%%
\subsection{LOFAR counterparts and spectral index}
\label{sec:SI}

Since the observing frequency of the LoTSS survey is 10 times lower than that of Apertif a combination of the two surveys provides unique information about the spectra of the sources, which, in turn, contain information about the emission mechanisms allowing to distinguish between the emitting regions of the sources (e.g., a flat spectrum radio cores vs optically thin jet components with steep spectra). Spectral index measurements can be used to estimate source ages~\citep[see][]{1962SvA.....6..317K}. 

\citet{2016MNRAS.460.2385W} and~\citet{2016MNRAS.463.2997M} used LOFAR images of the B\"ootes and Lockman Hole fields respectively to derive the spectral index distribution between 150 and 1400\,MHz. Their analysis results in very accurate median values of the spectral index of $-0.79 \pm 0.01$ and $-0.78 \pm 0.02$ correspondingly. These studies were limited to a relatively small area and the effect of different spatial resolution between the 150 and the 1400 MHz images could be taken into account. The authors used both resolved and unresolved sources for the spectral index estimation.

Because of the different resolution between the Apertif and LOFAR catalogs, we have limited our analysis to those sources unresolved in the latter, noting that these sources must be unresolved in the former as well. We have selected these sources following the criteria described by~\citet[Section 3.1, eq.~2]{2022A&A...659A...1S}, when the compactness depends on the ratio between integrated and peak flux density of a source as well as signal-to-noise level of its detection. Consistent with this, we find that about 90\% of the common sources are unresolved, which is close to the fraction of 95\% obtained for all LOFAR sources. 
We estimate the spectral index as $\alpha=\ln{(S_{\rm total,A}/S_{\rm total,L})}/\ln{(1355/150)}$, where $S_{\rm total,A}, S_{\rm total,L}$ are the Apertif and LOFAR total flux density of a source, and 1355\,MHz and 150\,MHz are the corresponding survey frequencies.
 
\begin{figure}
    \centering
    \includegraphics[width=\linewidth]{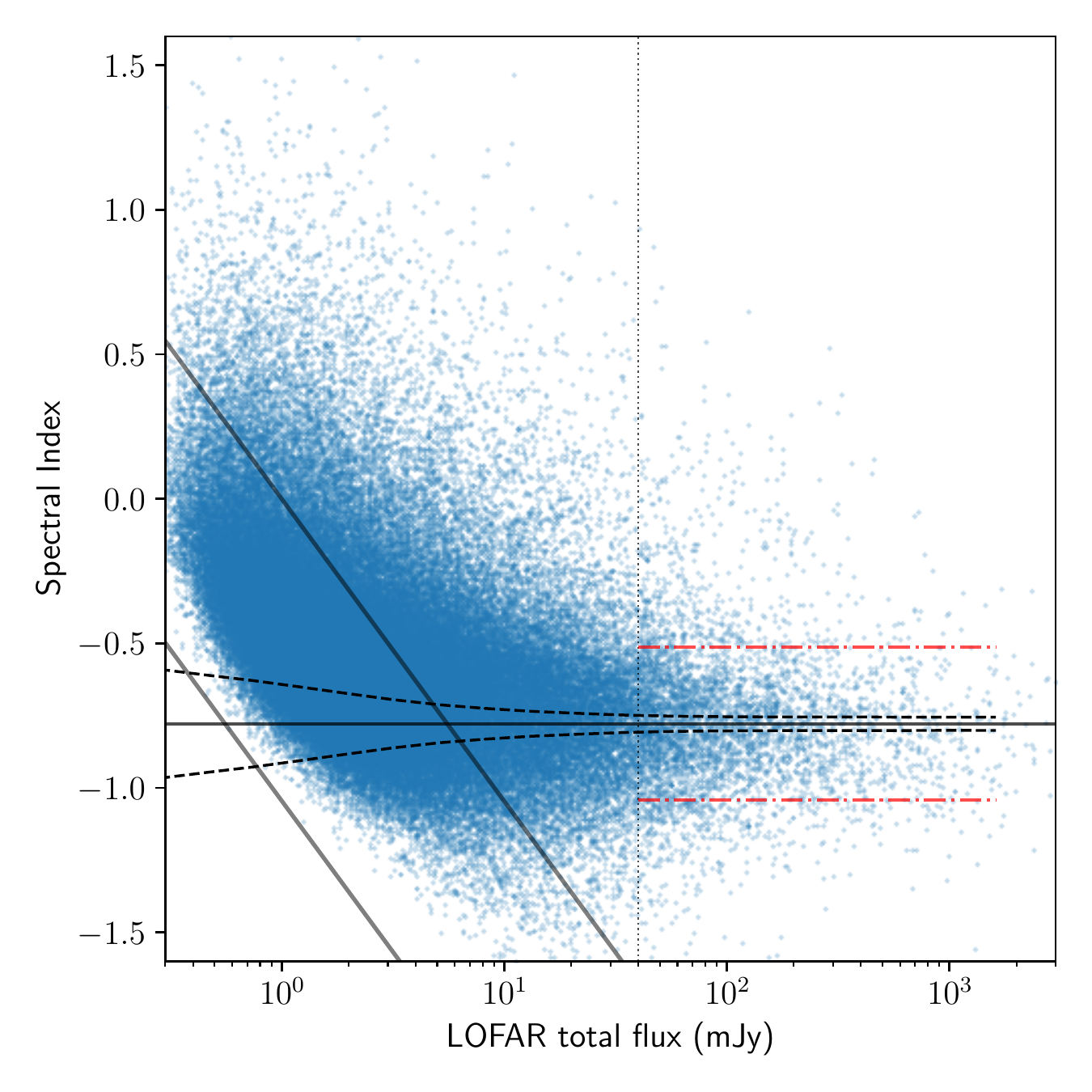}
    \caption{Spectral index distribution against the LOFAR integrated flux density for compact sources. The median value $\alpha=-0.78$ (horizontal line) is calculated for the sources with integrated flux density above \mjy{40} at 150\,MHz. The surrounding dashed curves show $\pm \sigma_{\alpha,S}$ error due to flux density uncertainties. The inclined solid lines show spectral index lower limit calculated for the weakest Apertif source and for \mjy{1} level corresponding to the completeness of 95\%. An overall scatter, $\pm \sigma_\alpha$, of the spectral index is shown with dash-dotted lines.}
    \label{fig:aplo_spindex}
\end{figure}

The spectral index distribution as a function of the LOFAR total flux is shown in Figure~\ref{fig:aplo_spindex}. The trend of flattening of the spectral index when going to lower fluxes occurs mostly due to the combination of increasing scatter for low flux sources and the Apertif sensitivity limit / completeness. This is illustrated by the overplotted curves. The envelope shown with the dashed line corresponds to the error due to the flux density uncertainties and is derived as follows:
$$
\sigma_{\alpha,S} = \frac{1}{\ln{(1355\,\mathrm{MHz}/150\,\mathrm{MHz})}}
\sqrt{
\left(\frac{\sigma_{S_\mathrm{total}}}{S_\mathrm{total}}\right)^2_{\mathrm{Apertif}} 
+
\left(\frac{\sigma_{S_\mathrm{total}}}{S_\mathrm{total}}\right)^2_{\mathrm{Lofar}}
}
$$
The left solid line shows the sensitivity limit calculated for the lowest Apertif peak flux, and the right one indicates the limit for the Apertif catalog completeness level of 1 mJy. Therefore, an estimate of a median spectral index is unbiased for the sources with LOFAR flux density above \mjy{40} (vertical dotted line). This estimate is $\alpha=-0.78$, with the scatter of $\sigma_\alpha = 0.26$ (horizontal dash-dotted lines). As seen in Figure~\ref{fig:aplo_spindex} most of the scatter comes from real differences in spectral index of the sources, while the uncertainty due to flux density measurements is relatively small.

In addition to the cross-matching with LoTSS\,DR2, we cross-match the Apertif catalog with the value-added LoTSS\,DR1 catalog. The latter has, among other parameters, optical counterparts and redshifts estimates for the sources~\citep{2019A&A...622A...3D}, which might be useful for studying either individual objects or statistical populations. The common area of the two catalogs is about 100 square degrees resulting in \aplova matched sources. 

We publish the list of common Apertif-LOFAR sources in Table~\ref{tab:aptf_lofar}. The table contains names (columns 1,2), peak and integrated flux density (columns 3--6) from the Apertif and LotSS/DR2 catalogs, spectral index estimate calculated for the peak fluxes (column 7), angular separation between the sources (column 8) and redshift estimate from LoTSS\,DR1 value-added catalog (column 9; \textit{zbest} as described by \citealt{2019A&A...622A...3D}). 
With this list we significantly increase the population of radio sources with spectral indexes measurements. The results can be used for studying, for example, individual objects, while the approach itself might be relevant for future deeper data releases. 

\begin{table*}
{\small \centering
    \caption{Common Apertif and LOFAR sources. \label{tab:aptf_lofar}}
    \begin{tabular}{llccccccc}
\toprule
        Apertif name &              Lofar name &    S$_\mathrm{peak}$$^\mathrm{A}$ &    S$_\mathrm{peak}$$^\mathrm{L}$ & S$_\mathrm{int}$$^\mathrm{A}$ & S$_\mathrm{int}$$^\mathrm{L}$ &     $\alpha$ &   $\phi$ & $z$ \\
                   &                       & [mJy/beam] & [mJy/beam] &  [mJy] &  [mJy] &      &   ["] & \\
\midrule
 APTF\_J132126+544301 &  ILTJ132126.92+544301.6 &       0.44 &       1.15 &   0.68 &   1.55 &  -0.37 &  1.87 &  0.79 \\
 APTF\_J132126+561608 &  ILTJ132126.56+561609.0 &       0.25 &       0.71 &   0.35 &   1.78 &  -0.74 &  1.99 &  0.99 \\
 APTF\_J132126+282552 &  ILTJ132126.86+282552.2 &       1.08 &       1.03 &   1.19 &   1.47 &  -0.10 &  0.27 &   \dots    \\
 APTF\_J132126+274020 &  ILTJ132126.95+274018.9 &       0.65 &       2.54 &   0.80 &   4.71 &  -0.81 &  1.20 &   \dots    \\
 APTF\_J132127+312037 &  ILTJ132126.90+312033.0 &       3.63 &      15.05 &   5.81 &  26.29 &  -0.69 &  4.49 &  \dots     \\
 APTF\_J132127+300946 &  ILTJ132127.21+300943.9 &       1.88 &       4.75 &   2.85 &  29.51 &  -1.06 &  2.21 &  \dots     \\
 APTF\_J132127+545520 &  ILTJ132127.22+545520.9 &       0.24 &       0.64 &   0.31 &   0.88 &  -0.47 &  1.19 &  0.53 \\
 APTF\_J132127+303538 &  ILTJ132127.46+303536.9 &       0.38 &       1.05 &   0.60 &   1.67 &  -0.47 &  2.40 &  \dots     \\
 APTF\_J132127+563932 &  ILTJ132127.42+563932.5 &       0.78 &       1.96 &   0.89 &   2.53 &  -0.47 &  0.62 &  0.32 \\
 APTF\_J132127+614427 &  ILTJ132127.25+614427.7 &       0.26 &       1.56 &   0.31 &   2.01 &  -0.86 &  1.10 &  \dots     \\
 APTF\_J132127+594632 &  ILTJ132127.50+594632.0 &       0.57 &       2.04 &   0.69 &   2.87 &  -0.65 &  0.85 &  \dots     \\
 APTF\_J132127+570045 &  ILTJ132127.80+570045.5 &       0.36 &       0.98 &   0.37 &   1.88 &  -0.74 &  1.24 &  \dots    \\
 APTF\_J132127+545419 &  ILTJ132127.83+545419.2 &       1.76 &       5.17 &   2.24 &  10.78 &  -0.71 &  0.60 &  0.65 \\
\bottomrule
\end{tabular}

    \tablefoot{Sample from the list of cross-matched Apertif (1355\,MHz; columns labeled with A) and LOFAR (150\,MHz; columns labeled with L) sources. Spectral index, angular separation between the sources and the redshifts are listed in the last three columns. The full table will be
    available online through CDS.}
}
\end{table*}

%%%%%%%%%%%%%%%%%%%%%%%%%%%%%%%%%%%%%%%%%%%%%%%%%%%%%%%%%%%%
\section{Summary}\label{sec:summary}

To deliver a ready-to-use source catalog for the scientific community, we process and analyze 3072 continuum images of the first data release of the Apertif survey. During this work we encounter and solve some challenging technical problems. 

To obtain a primary beam model we propose and apply a new method based on Gaussian process regression. This machine learning approach is a very natural choice because a primary beam shape and its peak are known to be smooth (Gaussian-like) and have a value of unity, which makes the choice of a kernel function and priors on hyperparameters straightforward. This method can be easily applied to other facilities and has several advantages compared to  analytical modeling, or to the drift scanning procedure. The most important  is the time economy (every drift scanning takes longer than 12 hours). The method provides a ready-to-apply correction model for every compound beam effectively removing a frequency- and an antenna- dependence. Instead of many parameters of an analytical approximation there are just a few hyperparameters entering the kernel function, which significantly simplifies  further analysis. Importantly, it becomes easy to obtain a primary beam model for any given date by taking into account only the observations around this date. This makes tracking  primary beam variations easier. Finally, all possible sources of attenuation bias are effectively removed by adjusting the flux scale to the one of the NVSS catalog. The primary beam models obtained with this method were published along with the data products for the first Apertif data release. 

In order to increase the sensitivity of the images, we create linear mosaics of the individual images taking into account the different size and orientation of restoring beam in each of them. The mosaic of all images of the first Apertif data release covers ~970 square degrees of sky. We present a new continuum source catalog for these data. The catalog contains \nsources radio sources, many of which are detected for the first time at L-band (1400\,MHz). The full catalog has a completeness of about 95\% at the \mjy{1} level. The number of false positive detections does not exceed a few percent. The table contains coordinates, integrated/peak flux density and angular sizes of the sources. 

We cross-matched the new Apertif catalog with the NVSS and the LOFAR/DR2 catalogs. On the one hand, this procedure provides an extra check for the obtained parameters of the sources, such as their fluxes and coordinate offsets. On the other hand, it has a large potential for a scientific research. The first sample provides an opportunity to detect long term transient sources which have significantly changed their flux density over the last 25 years. The second one includes flux measurements at 150 and 1400 MHz providing information about spectral properties of more than a hundred thousand sources. The redshift estimates adopted from the value-added LoTSS catalog might be useful for studying individual objects as well as volume limited samples. 

The released images were selected based on strong validation criteria providing a non-uniform sky coverage. We are, therefore, looking forward to applying the described methods and techniques to the future data releases, where the amount and quality of the images will be significantly increased.

%%%%%%%%%%%%%%%%%%%%%%%%%%%%%%%%%%%%%%%%%%%%%%%%%%%%%%%%%%%%
\begin{acknowledgements}

We thank the anonymous referee whose comments helped to improve the paper. 
This work makes use of data from the Apertif system installed at the Westerbork Synthesis Radio Telescope owned by ASTRON. ASTRON, the Netherlands Institute for Radio Astronomy, is an institute of the Dutch Science Organisation (De Nederlandse Organisatie voor Wetenschappelijk Onderzoek, NWO). Apertif was partly financed by the NWO Groot projects Apertif (175.010.2005.015) and Apropos (175.010.2009.012).
% Python + common
This research made use of \texttt{Python} programming language with its standard and external libraries/packages including \texttt{numpy}~\citep{numpy}, \texttt{scipy}~\citep{scipy}, \texttt{scikit-learn}~\citep{scikit-learn}, \texttt{matplotlib}~\citep{matplotlib}, \texttt{pandas}~\citep{pandas} etc.
% Astropy
This research made use of Astropy,\footnote{http://www.astropy.org} a community-developed core Python package for Astronomy \citep{astropy:2013, astropy:2018}. 
% RadioBeam/Reproject
The \texttt{radio\_beam} and \texttt{reproject} python packages are used for manipulations with restoring beam and reprojecting/mosaicing of the images. 
% Aladin 
This research has made use of "Aladin sky atlas" developed at CDS, Strasbourg Observatory, France~\citep{2000A&AS..143...33B, 2014ASPC..485..277B}. 
EAKA is supported by the WISE research programme, which is financed by the Netherlands Organization for Scientific Research (NWO)
BA acknowledges funding from the German Science Foundation DFG, within the Collaborative Research Center SFB1491 ''Cosmic Interacting Matters - From Source to Signal''
TAO acknowledge funding from  NWO via grant TOP1EW.14.105.
KMH acknowledges financial support from the State Agency for Research of the Spanish Ministry of Science, Innovation and Universities through the "Center of Excellence Severo Ochoa" awarded to the Instituto de Astrofísica de Andalucía (SEV-2017-0709), from the coordination of the participation in SKA-SPAIN, funded by the Ministry of Science and Innovation (MCIN). 
KMH and JMvdH acknowledge funding from the Europeaní Research Council under the European Union’s Seventh Framework Programme (FP/2007-2013)/ERC Grant Agreement No. 291531 (‘HIStoryNU’).
LC and LCO acknowledge funding from the European Research Council under the European Union's Seventh Framework Programme (FP/2007-2013)/ERC Grant Agreement No. 617199.
JvL acknowledges funding from Vici research programme `ARGO' with project number 639.043.815, financed by the Dutch Research Council (NWO).
DV acknowledges support from the Netherlands eScience Center (NLeSC) under grant ASDI.15.406

\end{acknowledgements}

%%%%%%%%%%%%%%%%%%%%%%%%%%%%%%%%%%%%%%%%%%%%%%%%%%%%%%%%%%%%%%%%%%%%%%

\appendix
\section{Errors estimation}
\label{sec:errors}

\subsection{Flux scale}
As shown in Figure~\ref{fig:flux_acc} even the bright sources have about 5\% uncertainty in their flux density measurements. This uncertainty likely comes from the cross-calibration procedure~(A22). In order to take this into account, we include the 5\% uncertainty into the final error of the flux density measurements in Table~1 by summing in quadrature $0.05S$ and the fitting error obtained from PyBDSF fit $\sigma_\mathrm{*,fit}$:
$$
\sigma_\mathrm{S,tot} = \sqrt{{(0.05S_\mathrm{tot}})^2 + \sigma_\mathrm{tot,fit}^2},\,
\sigma_\mathrm{S,peak} = \sqrt{{(0.05S_\mathrm{peak}})^2 + \sigma_\mathrm{peak,fit}^2}.
$$

\subsection{Coordinates}
The coordinate uncertainties of the LoTSS DR2 catalog are much smaller ($\sim0.2''$) than those of Apertif, which has lower resolution. Therefore, the former catalog can be used as a reference. Additionally to the fitting errors, $\sigma_\mathrm{RA,fit}$ and $\sigma_\mathrm{Dec,fit}$, provided by PyBDSF there are astrometric errors, $\sigma_\mathrm{RA,ast}$ and $\sigma_\mathrm{Dec,ast}$, coming from any systematic effects in an individual mosaic image. As shown in Figure~\ref{fig:aplo_coords} the coordinates offsets are scattered with a dispersion ranging from $\sim$\asec{2} to \asec{0.5} depending on source SNR. We fitted the percentile curves with hyperbola $\sigma_\mathrm{*,ast}=1/(a_*\cdot SNR+b_*)^{n_*}$, resulting in $a_\mathrm{RA}=0.031$, $b_\mathrm{RA}=-0.132$, $n_\mathrm{RA}=0.220$ and $a_\mathrm{Dec}=0.047$, $b_\mathrm{Dec}=-0.207$, $n_\mathrm{Dec}=0.281$. The fit was used to predict astrometric errors of the coordinates depending on SNR of a source. The final coordinate errors reported in Table~1 are then calculated as a quadratic sum of the astrometric error and the error from PyBDSF fitting:
$$
\sigma_\mathrm{RA} = \sqrt{\sigma_\mathrm{RA,ast}^2 + \sigma_\mathrm{RA,fit}^2},\,\,\,
\sigma_\mathrm{Dec} = \sqrt{\sigma_\mathrm{Dec,ast}^2 + \sigma_\mathrm{Dec,fit}^2}
$$
Average coordinate errors are $\overline{\sigma_\mathrm{RA}}=1.7''$ and $\overline{\sigma_\mathrm{Dec}}=2.1''$.

\subsection{Deconvolved size}

Along with the FWHM of a gaussian fitted to a source PyBDSF provides also the deconvolved source size and position angle (PA), and the corresponding errors. The latter, however, are not calculated correctly and remain the same as the errors of the original fit (the PyBDSF version used in this work is 1.9.2). To estimate the errors of a source size and PA after deconvolution, $\sigma_{\rm maj,min}$ and $\sigma_{\rm PA}$, we used the approach implemented in \texttt{CASA imfit}\footnote{\url{https://casa.nrao.edu/docs/taskref/imfit-task.html}}: 

\begin{quote}
    The deconvolved size and position angle errors are computed by taking the maximum of the absolute values of the differences of the best fit deconvolved value of the given parameter and the deconvolved size of the eight possible combinations of (FWHM major axis +/- major axis error), (FWHM minor axis +/- minor axis error), and (position angle +/- position angle error). If the source cannot be deconvolved from the beam (if the best fit convolved source size cannot be deconvolved from the beam), upper limits on the deconvolved source size are reported, if possible. These limits simply come from the maximum major and minor axes of the deconvolved Gaussians taken from trying all eight of the aforementioned combinations.
\end{quote}

In the case none of these combinations produces a deconvolved size, we estimate the upper limit using the major/minor axis FWHM of the restoring PSF and source SNR. A point source has a fitted size of the restoring PSF, and the corresponding error of the fitted size is $\sigma^{\rm fit}_{\rm maj, min} \approx {\rm PSF_{maj,min}/SNR}$~\cite[e.g.,][]{1999ASPC..180..301F}. Then the error after deconvolution can be estimated as the smallest deconvolved size detectable at $1\sigma$ level:

\begin{align*}
\sigma_{\rm maj,min} = (({\rm PSF_{maj,min} + PSF_{maj,min}/SNR})^2 - {\rm PSF^2_{maj,min}})^{1/2} = \\
{\rm \frac{PSF}{SNR}\left(1+2SNR\right)^{1/2}}.
\end{align*}

If both major and minor deconvolved size are zero, the position angle is omitted. The deconvolved source size, PA and the corresponding errors are listed in columns~10 through 15 of the Table~\ref{tab:catalog}.

%%%%%%%%%%%%%%%%%%%%%%%%%%%%%%%%%

\bibliographystyle{aa}
\bibliography{refs}

%%%%%%%%%%%%%%%%%%%%%%%%%%%%%%%%%%%%%%%%%%%

\end{document}